\documentclass[preprint,onecolumn,secnumarabic,nobalancelastpage,amsmath,superscriptaddress,amssymb,nofootinbib,achemso]{revtex4}
\usepackage{graphicx}
\usepackage{color}
\usepackage{chemmacros}
\usepackage{filecontents}
\usepackage[T1]{fontenc}
\usepackage[latin9]{inputenc}
\usepackage[latin9]{inputenc}
\usepackage{array}
\usepackage{multirow}

\begin{document}
\bibliographystyle{nature1}

\title{Protein Collapse is Encoded in the Folded State Architecture}
\author{Himadri S. Samanta}\affiliation{Department of Chemistry, University of Texas at Austin, TX 78712}
\author{Pavel I. Zhuravlev}\affiliation{Biophysics Program, Institute for Physical Science and Technology, University of Maryland, College Park, MD 20742}
\author{Michael Hinczewski} 
\affiliation{Department of Physics, Case Western Reserve University, OH 44106}
\author{Naoto Hori}\affiliation{Department of Chemistry, University of Texas at Austin, TX 78712}
\author{ Shaon Chakrabarti}\affiliation{Biophysics Program, Institute for Physical Science and Technology, University of Maryland, College Park, MD 20742}
\author{D. Thirumalai} 
\affiliation{Department of Chemistry, University of Texas at Austin, TX 78712}
\affiliation{Biophysics Program, Institute for Physical Science and Technology, University of Maryland, College Park, MD 20742}
\date{\today}

\begin{abstract}
Folded states of single domain globular proteins, the workhorses in cells, are compact with high packing density. It is known that the radius of gyration, $R_g$, of both the folded and unfolded (created by adding denaturants) states increase as $N^{\nu}$ where $N$ is the number of amino acids in the protein.   The values of  the celebrated Flory exponent $\nu$ are, respectively,  $\approx \frac{1}{3}$ and $\approx 0.6$ in the folded and unfolded states, which coincide with those found in homopolymers in poor  and good solvents, respectively.  However,  the extent of compaction of the unfolded state of a protein under low denaturant concentration, conditions favoring the formation of the folded state, is unknown. This problem which goes to the heart of how proteins fold, with implications for the evolution of foldable sequences, is unsolved.  We develop a theory based on polymer physics concepts that uses the contact map of proteins as input to quantitatively assess collapsibility of proteins. The model, which includes only two-body excluded volume interactions and attractive interactions reflecting the contact map, has only expanded and compact states.   Surprisingly, we find that although protein collapsibility is universal, the propensity to be compact depends on the protein architecture.  Application of the theory to over two thousand proteins shows that the extent of collapsibility depends not only on $N$ but also on the contact map reflecting the native fold structure.  A major prediction of the theory is that $\beta$-sheet proteins are far more collapsible than structures dominated by $\alpha$-helices.  The theory { and the accompanying simulations, validating the theoretical predictions,}  fully resolve the apparent controversy between conclusions reached using different experimental probes assessing the extent of compaction of a couple proteins. { As a by product, we show that the theory correctly predicts the scaling of the collapse temperature of homopolymers as a function of the number of monomers.} { By calculating the criterion for collapsibility as a function of protein length we provide} quantitative insights into the reasons why single domain proteins are small and the physical reasons for the origin of multi-domain proteins.   We also show that non-coding RNA molecules, whose collapsibility is similar to proteins with $\beta$-sheet structures, must undergo collapse prior to folding, adding support to ``Compactness Selection Hypothesis'' proposed in the context of RNA compaction.
\end{abstract}
\maketitle

\section{Introduction}

Folded states of globular proteins, which are evolved (slightly) branched heteropolymers made from twenty amino acids, are roughly spherical and are nearly maximally compact  with high packing densities \cite{Richards74JMB,Richards94QRB,Finney75JMB}. Despite achieving high packing densities in the folded states, globular proteins tolerate large volume substitutions while retaining the native fold~\cite{Lim89Nature}. This is explained in a couple of interesting theoretical studies~\cite{Liang01BJ,Dill94ProtSci}, which demonstrated that there is sufficient free volume in the folded state to accommodate mutations. Collectively these and related studies show that folded proteins are compact. When they unfold, which can be achieved upon addition of high concentrations of denaturants (or applying a mechanical force), they swell adopting  expanded conformations. The radius of gyration ($R_g$) of a folded globular protein is well described by the Flory law with $R_g \approx 3.3 N^{\frac{1}{3}}$ $\mathrm{\AA}$~\cite{Dima04JPCB}, whereas in the swollen state $R_g \approx a_D N^{\nu}$, where $a_D$ is an effective monomer size and the Flory exponent $\nu \approx 0.6$ \cite{Kohn04PNAS}. Thus, viewed from this perspective we could surmise that proteins must undergo a coil-to-globule transition \cite{Sherman06PNAS,Haran12COSB}, a process that is reminiscent of the well characterized  equilibrium collapse transition in homopolymers \cite{Lifshitz78RMP,GrosbergBook}. The latter is driven by the balance between conformational entropy  and intra-polymer interaction energy resulting in the collapsed globular state. The swollen state is realized in good solvents (interaction between monomer and solvents is favorable) whereas in the collapsed state monomer-monomer interactions are preferred. The coil-to-globule transition in large homopolymers is akin to a phase transition. The temperature at which the interactions between the monomers roughly balance monomer-solvent energetics is the $\theta$ temperature. By analogy, we may identify high (low) denaturant concentrations with good (poor) solvent for proteins. 

Despite the expected similarities between the equilibrium collapse transition in homopolymers and the compaction of proteins, it is still debated whether the unfolded states of proteins under folding conditions are more compact  compared to the states created at high denaturant concentrations. If polypeptide chain compaction is universal, is collapse in proteins  essentially the same phenomenon as in homopolymer collapse  or is it driven by a different mechanism \cite{Ptitsyn96NSB,Thirumalai95JPI,Ptitsyn94FEBSLett,Ding05JMB,Tran05Biochem}?  Surprisingly, this fundamental question in the protein folding field has not been answered satisfactorily~\cite{Ziv09PCCP,Haran12COSB}.   In order to explain the plausible difficulties in quantifying the extent of compaction, 
let us consider a protein, which undergoes an apparent two-state transition from an unfolded  (swollen) to a folded (compact) state as the denaturant concentration ($C$) is decreased. At the concentration, $C_m$, the populations of the folded and unfolded states are equal.  A vexing question, which has been difficult to unambiguously answer in experiments, is: what is the size, $R_g$,  of the unfolded state under folding conditions ($C < C_m$)? Small Angle X-ray Scattering (SAXS) experiments on some proteins show practically no change in the unfolded $R_g$ as $C$ is changed~\cite{Yoo12JMB}. On the other hand, from experiments based on single molecule Fluorescence Resonance Energy Transfer (smFRET) it has been concluded that the size of the unfolded state is more compact below $C_m$ compared to its value at high $C$~\cite{Schuler16JACS,Schuler08COSB}. The so-called smFRET-SAXS controversy is unresolved. Resolving this apparent controversy is not only important in our understanding of the physics of protein folding but also has implications for the physical basis of the evolution of  natural sequences. 


The difficulties in describing the collapse of unfolded states as $C$
is lowered could be attributed to the following reasons.  (1)
Following de Gennes~\cite{deGennes85JPF}, homopolymer collapse can be
pictured as formation of a large number of the blobs driven by local
interactions between monomers on the scale of the blob
size. Coarsening of blobs results in the equilibrium globule formation
with the number of maximally compact conformations whose number scales
exponentially with the number of monomers.   Other scenarios resulting in  fractal
globules, enroute to the formation of equilibrium maximally collapsed
structures, have also been proposed \cite{Grosberg88JdePhysique}. The globule formation is
driven by non-specific interactions between the monomers or the blobs.
Regardless of how the equilibrium globule 
is reached it is clear that it is largely stabilized by local interactions, because 
contacts between monomers that are distant along the sequence are entropically unfavorable. 
In contrast, even in high denaturant concentrations proteins could have
residual structure, which likely becomes prominent at $C < C_m$. At
low $C$ there are specific favorable interactions between residues
separated by a few or several residues along the sequence. As their
strength grows, with respect to the entropic forces, the specific
interactions may favor compaction in a manner different from the way
non-specific local interactions induce homopolymer collapse. In other
words, the dominant native-like contacts also drive compaction
of unfolded states of proteins. (2) A consequence of the impact of the
native-like contacts (local and non-local) on collapse of
unfolded states is that specific energetic considerations dictate
protein compaction resulting in the formation of minimum energy
compact structures (MECS)~\cite{Camacho93PRL}. The number of MECS, which
are not fully native, is small, scaling as
$\ln N$ with $N$ being the number of amino acid
residues. Therefore, below $C_m$ their contributions to $R_g$ have to
be carefully dissected, which is more easily done in single molecule
experiments than in ensemble measurements such as SAXS. (3) Single
domain proteins are finite-sized with $N$ rarely exceeding $\sim$
200. Most of those studied experimentally have $N <
  100$. Thus, the extent of change in $R_g$ of the unfolded states is
predicted to be small, requiring high precision
experiments to quantify the changes in $R_g$ as $C$ is changed. For
example, in a recent study \cite{Liu16JPCB}, we showed that in PDZ2
domain the change in $R_g$ of the unfolded states as the denaturant
concentration changes from 6 M guanidine chloride to 0 M is only about
8\%. Recent experiments have also established that changes in $R_g$ in
helical proteins are small \cite{Schuler16JACS}.

In homopolymers there are only two possible states, coil and globule,
with a transition between the two occurring at $T_{\theta}$. On the
other hand, even in proteins that fold in a two-state manner one can
conceive of at least three states (we ignore intermediates here):
(i) the unfolded state $\bf{U_D}$ at high $C$; (ii) the
  compact but unfolded state $\bf{U_C}$, which could possibly exist
  below $C_m$; (iii) the native state. 
Do the sizes of $\bf{U_D}$ and $\bf{U_C}$ differ? This question
requires a clear answer as it impacts our understanding of
how proteins fold, because the characteristics of the unfolded
states of proteins plays a key role in determining protein foldability
\cite{Camacho93PNAS,Li04PRL,Dill91AnnRevBiochem}.


Given the flexibility of proteins (persistence length on the order of
$0.5 - 0.6$ nm), we expect that the size of the extended
polypeptide chain must gradually decrease as the solvent quality is
altered. Experiments on a number of proteins show that this is the
case~\cite{Akiyama02PNAS,Kimura05JMB,Goluguri16JMB}. However, in some
SAXS experiments the theoretical expectation that
$R_g^{\bf{U_C}} < R_g^{\bf{U_D}}$ for one protein was not borne out
\cite{Yoo12JMB,Haran12COSB}, precipitating a more general
question: are chemically denatured proteins compact at low
$C$? The absence of collapse is not compatible with inferences based
on smFRET \cite{Schuler08COSB} and theory \cite{Camacho93PNAS}.  Here,
we create a theory to not only resolve the smFRET-SAXS controversy but
also provide a quantitative description of how the propensity to be
compact is encoded in the native topology. The theory, based on
polymer physics concepts, includes specific attractive interactions
(mimicking interactions accounting for native contacts in the Protein
Data Bank (PDB)) and a two-body excluded volume repulsion.  By
construction the model does not have a native state. {In order to validate the theoretical predictions, we performed simulations using a completely different model often used in protein folding simulations.} In both the models, there are only
two states (analogues of $\bf{U_D}$ and $\bf{U_C}$) in the model. The
formation of $\bf{U_C}$ is driven by the contact map of the folded
state. Thus, chain compaction is driven in much the same way as in
homopolymers, altered only by specific interactions that differentiate
proteins from homopolymers. 

{ Theory and simulations} predict how the extent of compaction
(collapsibility) is determined by the strength and the number of the
native contacts and their locations along the chain. We use a large
representative selection of proteins from the PDB to
establish that collapsibility is an inherent characteristic of evolved
protein sequences.  A major outcome of this work is that $\beta$-sheet
proteins are far more collapsible than structures dominated by
$\alpha$-helices.  Our theory suggests that there is an evolutionary
pressure on proteins for being compact as a pre-requisite for kinetic
foldability, as we predicted over twenty years ago
\cite{Camacho93PNAS}. We come to the inevitable conclusion that the
unfolded state of proteins must be compact under native conditions,
and the mechanism of polypeptide chain compaction has
similarities as well as differences to collapse in
homopolymers. As a by-product of this work, we also
establish that certain non-coding RNA molecules must undergo
compaction prior to folding as their folded structures are stabilized predominantly
by long-range tertiary contacts.

\section{Theory}
We start with an Edwards Hamiltonian for a polymer chain~\cite{Edwards65PRC}:
\begin{equation}\label{hamiltonian}
\mathcal{H}=\frac{3k_B T}{2 a_0^2} \int\limits_0^N  \left(\frac{\partial {\bf r}}{\partial s}\right)^2 ds + k_B T  \mathcal{V}({\bf r}(s)),
\end{equation}
where ${\bf r}(s)$ is the position of the monomer $s$, $a_0$ the monomer size, and $N$ is the number of monomers. The first term in Eq.~(\ref{hamiltonian}) accounts for chain connectivity, and the second term represents volume interactions and favorable interactions between select monomers given by $\mathcal{V}({\bf r}(s))$,

\begin{equation}\label{HamiltV}
\mathcal{V}({\bf r}(s))=\frac{v}{(2\pi a_0^2)^{3/2}}\sum\limits_{s=0}^{N}  \sum\limits_{s'=0}^{N}  
e^{-\frac{({\bf r}(s)-{\bf r}(s'))^2}{2a_0		^2}}-
\frac{\kappa}{(2\pi \sigma^2)^{3/2}}\sum\limits_{\{s_i,s_j\}} e^{-\frac{({\bf r}(s_{i})-{\bf r}(s_{j}))^2}{2\sigma^2}}
\end{equation}

The first term in Eq.(\ref{HamiltV}) accounts for  the homopolymer (non-specific) two-body interactions.
It is well established in the theory of homopolymers that  in good solvents with $v > 0$ the polymer swells with $R_g \sim a N^{\nu}$ $(\nu \approx 0.6)$. In poor solvents ($v < 0$)  the polymer undergoes a coil-globule transition with $R_g \sim a N^\nu$ ($\nu \approx 1/3$). These are the celebrated Flory laws. Here, we consider only the excluded volume repulsion case ($v>0$).

The second term in Eq.~(\ref{HamiltV}) requires an explanation. The
generic scenario for homopolymer collapse is based on an
observation by de Gennes, who pictured the collapse
process as being driven by the initial formation of blobs that arrange
to form a sausage-like structure. At later stages the globule forms 
to maximize favorable intra-molecular contacts while
simultaneously minimizing surface tension.
Compaction in proteins, although shares many features in
common with homopolymer collapse, could be different.  A key
difference is that the folded states of almost all proteins are
stabilized by a mixture of local contacts (interaction between
residues separated by less than say $\sim$ 8 but greater than 3
residues) as well as non-local ($>$ 8 residues) contacts.  Note that
the demarcation using 8 between local and non-local contacts is
arbitrary, and is not germane to the present argument.
These specific interactions also dominate the enthalpy of
  formation of the compact, non-native state $\bf{U_C}$, playing an
  important role in its stability.  { Previous studies using lattice models of proteins in two \cite{Camacho95Proteins} and three \cite{Klimov01Proteins} dimensions showed that formation of compact but unfolded states are predominantly driven by native interactions with non-native interactions playing a sub-dominant role. A more recent study \cite{Best13PNAS}, analyzing atomic detailed folding trajectories has arrived at the same conclusion.} Therefore, { our assumption is that} the topology of the
folded state could dictate collapsibility (the extent to which the
$\bf{U_D}$ state becomes compact as the denaturant concentration is
lowered) of a given protein. In combination with the
finite size of single domain proteins ($N$ $\sim$ 200),
the extent of protein collapse could be small. In order to assess
chain compaction under native conditions we should consider the second term
in Eq.(\ref{HamiltV}).

{ It is worth mentioning that several studies investigated the consequences of optimal packing of polymer-like representations of proteins ~\cite{Yasuda10JCP, Maritan00Nature, Magee06PRL, Skrbic16JCP, Snir05Science, Craig06Macromolecules, Cardelli16arxiv}. These studies primarily explain the emergence of secondary structural elements by considering only hard core interactions, attractive interactions due to crowding effects \cite{Snir05Science,Kudlay09PRL}, or formation of compact states induced by anisotropic attractive patchy interactions~\cite{Cardelli16arxiv}. However, the absence of tertiary interactions in these models, which give rise to compact states of varying topologies, prevents them from addressing the coil-to-globule transition. This requires creating a microscopic model along the lines described here.}

We note in passing (with discussion to follow) that a number of studies have considered the effect of crosslinks on the shape of polymer chains \cite{Gutin94JCP,Bryngelson96PRL,Solf96PRL,Kantor96PRL,Zwanzing97JCP,Camacho96EPL,Camacho97EPL}. Polymers with crosslinks have served as models for polymer gels and rubber elasticity \cite{Deam76,GoldbartPRL87,GoldbartPRA89}. In these studies the contacts were either random, leading to the random loop model \cite{Bryngelson96PRL}, or explicit averages over the probability of realizing such contacts were made \cite{Gutin94JCP,Catillo94EPL}, as may be appropriate in modeling gels. These studies inevitably predict a coil-to-globule phase transition as the number of crosslinks increases.

In contrast to models with random crosslinks, in our theory attraction
exists only between specific residues, described by the second term in
Eq.~(\ref{HamiltV}), where the sum is over the set of interactions
(native contacts) involving pairs $\{s_i,s_j\}$.  We use the contact map of the
protein (extracted from the PDB structure) in order to assign the
specific interactions (their total number being $N_\mathrm{nc}$). The
contact is assigned to any two residues $s_i$ and $s_j$ if the
distance between their $C_\alpha$ atoms in the PDB entry is less than
$R_c=0.8 nm$ and $|s_i-s_j|>2$.  We use Gaussian potentials in order to
have short (but finite) range attractive interactions. For the
excluded volume repulsion, this range is on the order of the size of the monomer,
$a_0=0.38$ nm. For the specific attraction, the range is the average
distance in the PDB entry between $C_\alpha$ atoms forming a contact
(averaged across a selection of proteins from the PBD). We obtain
$\sigma=0.63$ nm. 

By changing the value of $\kappa$, and hence the strength of
attraction, there is a transition between the extended and compact
states. Decreasing $\kappa$ is analogous to chemically denaturing
proteins, although the connection is not precise. At high
denaturant concentrations ($\kappa \approx 0$, good solvent) the
excluded volume repulsion (first term in Eq.(\ref{HamiltV})) dominates
the attraction, while at low $C$ (high $\kappa$, poor solvent) the
attractive interactions are important. The point where attraction
balances repulsion is the $\theta$-point, and the value of
$\kappa=\kappa_{\theta}$.  Although reserved for the
  coil-to-globule transition in the limit of $N \gg 1$ in
homopolymers, we will use the same notation ($\theta$-point)
here. In our model, at the
$\theta $-point, the chain behaves like an ideal chain. To
describe the globular state, a three-body
repulsion needs to be added to the Hamiltonian (Eq.~(\ref{HamiltV})), but we focus on the
region between the extended coil and the $\theta$-point because our
interest is to access only the collapsibility of proteins. If
$\kappa_{\theta}$ is very large then significant chain compaction
would only occur at very low ($C \ll C_m$) denaturant concentrations,
implying low propensity to collapse. Conversely, small
$\kappa_{\theta}$ implies ease of collapsibility. Note that the ground
state ($\kappa \gg 1$) of the Hamiltonian in Eq.~(\ref{HamiltV})
 is a collapsed chain whose $R_g$ is on the order
of the monomer size. In other words,
a stable native state does not exist for the model described in
Eq.~(\ref{HamiltV}). Thus, we define protein collapse as the propensity of the
polypeptide chain to reach the $\theta$-point as measured
by the $\kappa_{\theta}$ value, and use the changes in the radius of
gyration $R_{g}$ as a measure of the extent of compaction.

{\bf Assessing collapsibility:} For our model, which encodes protein topology without favoring the folded state, we calculate $\langle R_g^2 \rangle$ using the Edwards-Singh (ES) method \cite{Edwards:1979vb}. {  Although from a technical view point the ES method has pros as well as cons, numerous applications show that in practice it yields physically sensible results on a number of systems.  First, ES showed that the method does give the correct dependence of $\langle R_g^2 \rangle$ on $N$ for homopolymers. Second, even when attractive interactions are included,  the ES method leads to predictions, which have been subsequently verified by more sophisticated theories.  An example of particular relevance here is the problem of the size of a polymer in the presence of obstacles (crowding particles). The results of the ES method \cite{Thirumalai88PRA} and those obtained using renormalization group calculations \cite{Duplantier88PRA}  are qualitatively similar. Here, we adopt the ES method, allowing us to deduce far reaching conclusions for protein collapsibility than is possible solely based on simulations. We use simulations on a limited set of proteins to further justify the conclusions reached using the analytic theory.}  

The ES method is a variational type calculation that represents the exact Hamiltonian by a Gaussian chain, whose effective monomer size is determined as follows.
Consider a virtual chain without excluded volume interactions, with the radius of gyration $\langle R_{g}^{2} \rangle=N a^{2}/6$~\cite{Edwards:1979vb}, described by the Hamiltonian,
\begin{equation}
\mathcal{H}_v=\frac{3k_B T}{2 a^2} \int\limits_0^N  \left(\frac{\partial {\bf r}}{\partial s}\right)^2 ds,
\end{equation}
where the monomer size in the virtual Hamiltonian is $a$.
We split the deviation $\mathcal{W}$ between the virtual chain Hamiltonian and the real Hamiltonian as,
\begin{equation}
\mathcal{H}-\mathcal{H}_v=k_BT\mathcal{W}=k_BT(\mathcal{W}_1+\mathcal{W}_2),
\end{equation}
where
\begin{eqnarray}
\mathcal{W}_1&=&\frac{3}{2 }\left(\frac{1}{a_0^2}-\frac{1}{a^2}\right) \int\limits_0^N  \left(\frac{\partial {\bf r}}{\partial s}\right)^2 ds,  \nonumber \\
\mathcal{W}_2&=&\mathcal{V}({\bf r}(s)).
\end{eqnarray}
The radius of gyration is $R_g^2=\frac{1}{N} \int\limits_0^N \langle{\bf r}^2(s)\rangle ds$, with the average being,
\begin{equation}
	\langle{\bf r}^2(s)\rangle=\frac{\int r^2 e^{-\mathcal{H}/k_BT} \delta{\bf r}}{\int e^{-\mathcal{H}/k_BT} \delta{\bf r}}=\frac{\int r^2 e^{-\mathcal{H}_v/k_BT}e^{\mathcal{-W}} \delta{\bf r}}{\int e^{-\mathcal{H}_v/k_BT}e^{\mathcal{-W}} \delta{\bf r}}=\frac{\langle{\bf r}^2(s)e^{\mathcal{-W}}\rangle_v}{\langle e^{\mathcal{-W}}\rangle_v}
\end{equation}
where $\langle \cdots \rangle_v$ denotes the average over $\mathcal{H}_v$.

Assuming that the deviation $\mathcal{W}$ is small, we calculate the average to first order in $\mathcal{W}$. The result is, 
\begin{equation}
	\langle{\bf r}^2(s)\rangle \approx \frac{\langle{\bf r}^2(s)(1-\mathcal{W})\rangle_v}{\langle (1-\mathcal{W})\rangle_v} \approx \langle{\bf r}^2(s)(1-\mathcal{W})\rangle_v\langle (1+\mathcal{W})\rangle_v 
\end{equation}
and the radius of gyration is
\begin{equation}\label{rg}
\langle R_g^2\rangle=\frac{1}{N} \int\limits_0^N \langle{\bf r}^2(s)\rangle ds = \frac{1}{N} \int\limits_0^N [\langle{\bf r}^2(s)\rangle_v + \langle{\bf r}^2(s)\rangle_v \langle\mathcal{W}\rangle_v -\langle{\bf r}^2(s)\mathcal{W}\rangle_v] ds,
\end{equation}

If we choose the effective monomer size $a$ in $\mathcal{H}_v$ such that the first order correction (second and third terms on the right hand side of Eq.~(\ref{rg})) vanishes, then the size of the chain is, $\langle R_{g}^{2} \rangle=N a^{2}/6$. This is an estimate to the exact $\langle R_g^2 \rangle$, and is  an approximation as we have neglected $\mathcal{W}^2$ and higher powers of $\mathcal{W}$. Thus, in the ES theory, the optimal value of $a$ from Eq. (\ref{rg}) satisfies,
\begin{equation}\label{first}
 \frac{1}{N} \int\limits_0^N [ \langle{\bf r}^2(s)\rangle_v \langle\mathcal{W}\rangle_v -\langle{\bf r}^2(s)\mathcal{W}\rangle_v] ds=0.
\end{equation}
Since $\mathcal{W}=\mathcal{W}_1+\mathcal{W}_2$, the above equation can be written as
\begin{equation}
 \frac{1}{N} \int\limits_0^N [ \langle{\bf r}^2(s)\rangle_v \langle\mathcal{W}_1\rangle_v -\langle{\bf r}^2(s)\mathcal{W}_1\rangle_v] ds=-\frac{1}{N} \int\limits_0^N [ \langle{\bf r}^2(s)\rangle_v \langle\mathcal{W}_2\rangle_v -\langle{\bf r}^2(s)\mathcal{W}_2\rangle_v] ds.
\end{equation}
Evaluation of the $\langle{\bf r}^2(s)\mathcal{W}_1\rangle_v$ term yields,
\begin{eqnarray}\label{aa}
&&\langle{\bf r}^2(s)\mathcal{W}_1\rangle_v =
\frac{\frac{3}{2 }\left(\frac{1}{a_0^2}-\frac{1}{a^2}\right) \int {\bf r}^2 \int\limits_0^N  \dot{{\bf r}}^2 ds \   \ e^{-\frac{3}{2a^2}\int\limits_0^N \dot{{\bf r}}^2 ds} \delta{\bf r}}{\int e^{-\frac{3}{2a^2}\int\limits_0^N \dot{{\bf r}}^2 ds} \delta {\bf r}}
\\ \nonumber 
&=& \frac{3}{2 }\left(\frac{1}{a_0^2}-\frac{1}{a^2}\right)
\left [ \frac{\partial}{\partial \alpha}\left.\left(
\frac{\int \delta{\bf r} {\bf r}^2 e^{\alpha \int \dot{{\bf r}}ds}}{\int \delta{\bf r}  e^{\alpha \int \dot{{\bf r}}ds}}\right)\right|_{ \alpha=-\frac{3}{2a^2}}+
\frac{\int {\bf r}^2  \   \ e^{-\frac{3}{2a^2}\int\limits_0^N \dot{{\bf r}}^2 ds} \delta{\bf r}}{(\int e^{-\frac{3}{2a^2}\int\limits_0^N \dot{{\bf r}}^2 ds} \delta{\bf r})^2}\int  \int\limits_0^N  \dot{{\bf r}}^2 ds \   \ e^{-\frac{3}{2a^2}\int\limits_0^N \dot{{\bf r}}^2 ds} \delta{\bf r}
  \right ] \\ \nonumber
&=&\left(\frac{1}{a_0^2}-\frac{1}{a^2}\right) a^2 \left(\frac{a^2 N}{6} \right)+ \langle{\bf r}^2(s)\rangle_v \langle\mathcal{W}_1\rangle_v
\end{eqnarray}

With the help of Eq.~(\ref{aa}) and Eq.~(\ref{first}) we obtain the following self-consistent expression for $a$,
\begin{equation}
 \frac{1}{a_0^2}-\frac{1}{a^2}=
\frac{\frac{1}{N}\int_0^N[\langle{\bf r}^2(s)\rangle_v \langle\mathcal{V}\rangle_v -\langle{\bf r}^2(s)\mathcal{V}\rangle_v]ds}{ \frac{a^2}{N}\int_{0}^N ds \ \langle{\bf r}^2(s)\rangle_v}.
\end{equation}
Calculating the averages in Fourier space, where $\tilde{{\bf r}}_n=\frac{1}{N}\int\limits_1^N \cos\left({ \frac{\pi n s}{N}}\right) {\bf r}(s) ds$, ${\bf r}(s)=2\sum\limits_{n =1}^{N}\cos\left({\frac{\pi n s}{N}}\right)\tilde{{\bf r}}_n$, and $R_g^2=2\sum\limits_n \langle|{\tilde{{\bf r}}_n}^2|\rangle$), we obtain
\begin{eqnarray}\label{self}
  \frac{1}{a_0^2}&-&\frac{1}{a^2}=v\frac{ (\frac{3}{2})^{5/2}(\frac{\pi}{2})^{3/2}}{ (a^2)^{5/2}N^{3/2}\left(\sum\limits_{n=1}^N \frac{1}{n^2}\right)}
  \sum\limits_{s=0}^N  \sum\limits_{s'=0}^N   \frac{\sum\limits_{n=1}^{N}\frac{1-\cos[n \pi(s-s')/N]}{n^4}}
{\left(\sum\limits_{n=1}^{N}\frac{1-\cos[n \pi(s-s')/N]}{n^2}+\frac{3 \pi^2a_0^2 }{2a^2 N}\right)^{5/2}}\\ \nonumber
  &-& \kappa\frac{ (\frac{3}{2})^{5/2}(\frac{\pi}{2})^{3/2}}{(a^2)^{5/2}N^{3/2}\left(\sum\limits_{n=1}^N \frac{1}{n^2}\right)}
 \sum\limits_{\{s_i,s_j\}}  \frac{\sum\limits_{n=1}^{N}\frac{1-\cos[n \pi(s_i-s_j)/N]}{n^4}}
{\left(\sum\limits_{n=1}^{N}\frac{1-\cos[n \pi(s_i-s_j)/N]}{n^2}+\frac{3 \pi^2\sigma^2 }{2a^2 N}\right)^{5/2}}.\\ \nonumber
\end{eqnarray}

The best estimate of the effective monomer size $a$ can be obtained by numerically solving Eq.~(\ref{self}) provided the contact map is known. A bound for the actual size of the  chain is $\langle R_g^2 \rangle =Na_0^2/6$. Because we are interested only in the collapsibility of proteins  we use the definition of the $\theta$-point to assess the condition for protein compaction instead of solving the complicated Eq.~(\ref{self}) numerically. The volume interactions are on the right hand side of Eq.~(\ref{self}). At the $\theta$-point, the $v$-term  should exactly balance the $\kappa$-term. Since at the $\theta$-point the chain is ideal with $a=a_0$, we can substitute this value for $a$ in the sums in the denominators of the $v$- and $\kappa$-terms. By equating the two, we obtain an expression for $\kappa_\theta$.
Thus, from Eq.~(\ref{self}), the specific interaction strength at which two-body repulsion ($v$-term) equals two-body attraction ($\kappa$-term) is:
\begin{equation}
	\kappa_\theta=\frac{4}{3}\pi a_0^3\frac{\sum\limits_{s=0}^N  \sum\limits_{s'=0}^N   \frac{\sum\limits_{n=1}^{N}\frac{1-\cos[n \pi(s-s')/N]}{n^4}}
	{\left(\sum\limits_{n=1}^{N}\frac{1-\cos[n \pi(s-s')/N]}{n^2}+\frac{3 \pi^2 }{2 N}\right)^{5/2}}}{ \sum\limits_{\{s_i,s_j\}}  \frac{\sum\limits_{n=1}^{N}\frac{1-\cos[n \pi(s_i-s_j)/N]}{n^4}}
	{\left(\sum\limits_{n=1}^{N}\frac{1-\cos[n \pi(s_i-s_j)/N]}{n^2}+\frac{3 \pi^2\sigma^2 }{2a_0^2 N}\right)^{5/2}}}.
\label{ktheta}
\end{equation}
The numerator in Eq.~(\ref{ktheta}) is a consequence of chain connectivity and the denominator encodes protein topology through the contact map, determining the extent to which the sizes in $\bf{U_D}$ and $\bf{U_C}$ states change as $C$ becomes less than $C_m$. The numerical value of $\kappa_\theta$ is a measure of collapsibility. 

{ A comment about the solution of Eq.~(\ref{self}) for $a$ is worth making. For $\kappa = 0$, corresponding to the good solvent condition, we expect that $a \gg a_0$. In this case, analysis of Eq.~(\ref{self}), in a manner described in Appendix A, shows that there is only one solution with $a \sim N^{\frac{1}{10}}$. Similarly, at $k_{\theta}$ Eq.~(\ref{self}) also admits only one solution. Thus, from the structure of Eq. (\ref{self}) we surmise there are no multiple solutions, at least in the extreme limits $v=0$ and $k=0$.

The expression for $k_{\theta} $(Eq.~(\ref{ktheta})) is equally applicable to homopolymers in which contacts between all monomers are allowed, provided the self-avoidance condition is not violated. In Appendix A, we derive an expression for $k_{\theta} \propto T_{\theta} \sim v(1- (vN^{-0.5})/2)$. Thus, our model correctly reproduces the known $N$ dependence of $T_{\theta}$ obtained long ago by Flory~\cite{Flory} using insightful mean field arguments.}

\section{Results}
{\bf Native topology determines collapsibility:} The central result in Eq.~(\ref{ktheta}) can be used to quantitatively predict the extent to which a given protein has a propensity to collapse.
We used a list of proteins with low mutual sequence identity selected from the Protein Data Bank PDBselect~\cite{griep2009pdbselect}, and calculated $\kappa_\theta$ using Eq.~(\ref{ktheta}) for these proteins. In all we considered 2306 proteins. For each contact $(i,j)$, the energetic contribution due to interaction between $i$ and $j$ is $k=(2\pi\sigma^2)^{-3/2}\kappa$ according to Eq.~(\ref{HamiltV}). Thus, $k_\theta=(2\pi\sigma^2)^{-3/2}\kappa_\theta$ is the average strength (in units of $k_B T$) of a contact at the $\theta$-point.
If $\kappa_\theta$, calculated using Eq.~(\ref{ktheta}), is too large then the extent of polypeptide chain collapse is expected to be small.    It is worth reiterating that the theory cannot be used to determine the stability of the folded state, because in the Hamiltonian there are only two states, $\bf{U_D}$ ($\kappa$=0 in Eq.(\ref{HamiltV})) and $\bf{U_C}$ ($\kappa > \kappa_\theta$). 

The strength of contacts in real proteins (excluding possibly salt bridges) is typically on the order of a few $k_BT$ in the absence of denaturants. This is the upper bound for the contact strength any theory should predict, as adding denaturant only decreases the strength. If $k_\theta$ is unrealistically high (tens of $k_BT$) then the attractive interactions of the protein would be too weak to counteract the excluded volume repulsion even at zero denaturant concentration, resulting in negligible difference in $R_g$ between the $\bf{U_D}$ and $\bf{U_C}$ states.

\begin{figure}
	\includegraphics[width=0.6\textwidth]{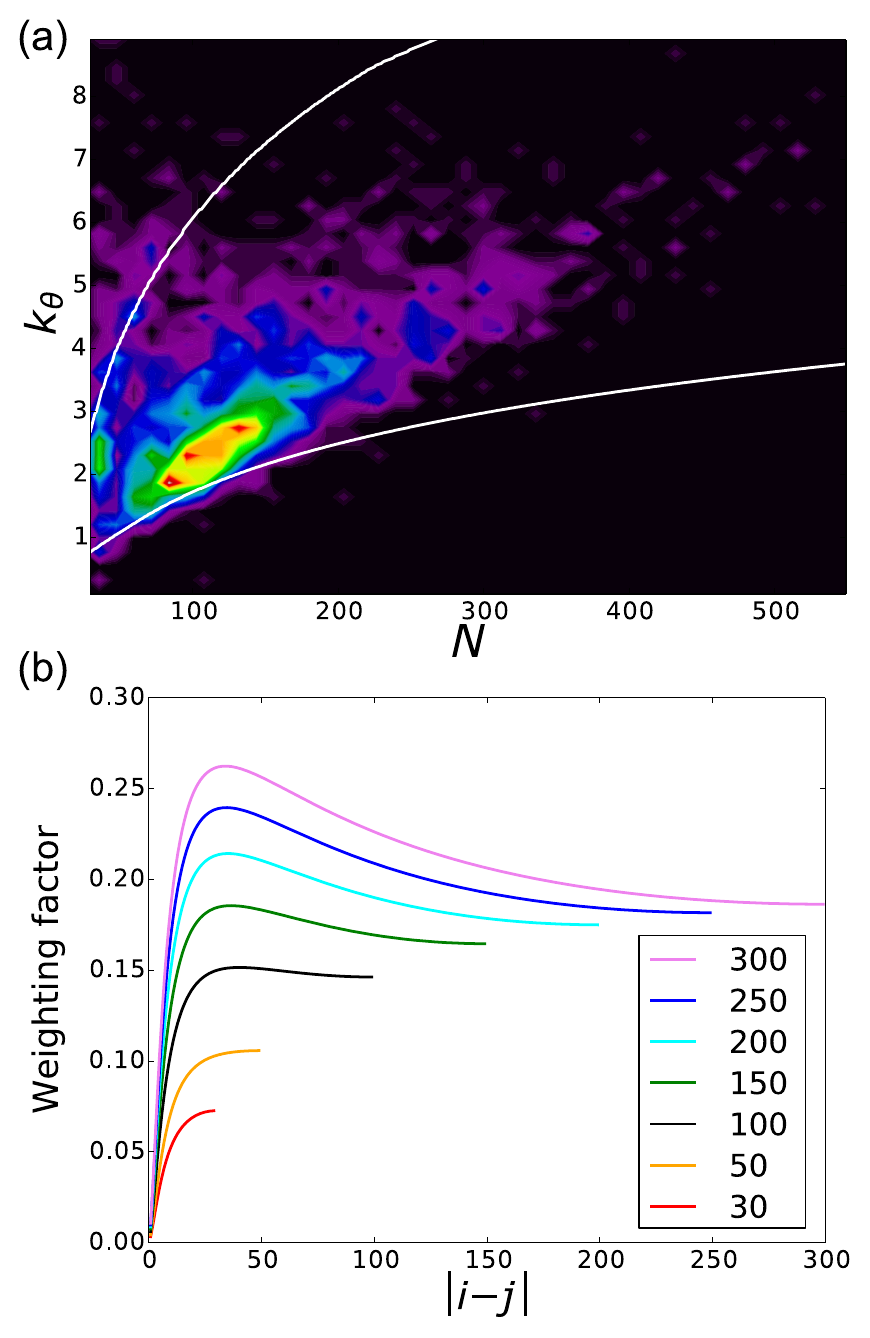}
	\caption{(a) Collapsibility quantified using $k_\theta$ (in units of $k_BT$) for a set of 2,306 PDB structures as a function of the length $N$ of the proteins. White lines show the $k_\theta$ at the boundaries for maximally and minimally collapsible proteins (lower and upper lines respectively). Colors give a rough estimate of the number of proteins, which decreases from red to violet. A dynamic visualization of the data is available at author's website~\cite{collapseURL}. (b) Weight function $W$ (Eq.(\ref{weighting})) of a contact, showing how much a contact between residues $i$ and $j$ contributes to the compaction of a protein. The colors are for different $N$ values (shown in the inset). Interestingly, the location of the maximum is roughly independent of $N$. }
		\label{fig:pdbhist}
\end{figure}

Fig.(\ref{fig:pdbhist}a) shows a two-dimensional histogram of the
PDBselect proteins in the $(N,k_\theta)$ plane. For the
majority of small proteins (less than 150 residues) the value of
$\kappa_\theta$ is less than 3 $k_BT$, indicating that the unfolded
states of all of these proteins should become compact at $C<C_m$. That
collapse must occur, as predicted by our theory and established
previously in lattice~\cite{Camacho93PNAS}, and off-lattice models of
proteins\cite{HoneycuttBP92}, does not necessarily imply that it can
be easily detected in standard scattering experiments, because the
changes could be small requiring high precision experiments (see
below).

{\bf Weight function of a contact:} For a given $N$, the criterion for collapsibility in Eq.~(\ref{ktheta}) depends on the architecture of the proteins explicitly represented in the denominator through the contact map. Analysis of the weight function of a contact, defined below, provides a quantitative measure of how a specific contact influences protein compaction. Some contacts may facilitate collapse to a greater extent than others, depending on the location of the pair of residues in the polypeptide chain. In this case, the same number of native contacts $N_\mathrm{nc}$ in the protein of the same length $N$ might yield a lower (easier collapse) or higher (harder collapse) value of $k_\theta$. In order to determine the relative importance of the contacts with respect to  collapse, we consider the contribution of the contact between residues $i$ and $j$ in the denominator of Eq.~(\ref{ktheta}),
\begin{equation}
	W(i-j)=\frac{\sum\limits_{n=1}^{N}\frac{1-\cos[n \pi(i-j)/N]}{n^4}}
	{\left(\sum\limits_{n=1}^{N}\frac{1-\cos[n \pi(i-j)/N]}{n^2}+\frac{3 \pi^2\sigma^2 }{2a_0^2 N}\right)^{5/2}}.
	\label{weighting}
\end{equation}
A plot of $W(i-j)$ in Fig.(\ref{fig:pdbhist}b) for different values
of the chain length $N$ shows that the weight depends on the distance
between the residues along the chain. Contacts between neighboring
residues have negligible weight, and there is a maximum in $W(i-j)$ at
$i-j\approx30$ (for $a_0/\sigma=0.6$),
almost independent of the protein length. The maximum is at a higher
value for proteins with $N>100$ residues. The figure further shows
that longer range contacts make greater contribution to chain
compaction than short range contacts. The results in
Fig.(\ref{fig:pdbhist}b) imply that proteins with a large
fraction of non-local contacts are more easily collapsible than those
dominated by short range contacts, which we elaborate further below.

{\bf Maximum and minimum collapsibility boundaries:} Using $W(i-j)$ in Eq.~(\ref{weighting}), we can design protein sequences to optimize for ``collapsibility''. To design a ``maximally collapsible'' protein, for fixed $N$ and number of native contacts $N_\mathrm{nc}$, we assign each of the $N_\mathrm{nc}$ contacts one by one to the pair $i,j$ with a maximal $W(i,j)$ among the available pairs with the criterion that $|i-j|>2$. Such an assignment necessarily implies that the artificially designed contact map will not correspond to any known protein. Similarly, we can design an artificial contact map by selecting $i,j$ pairs with minimal $W(i,j)$ till all the $N_\mathrm{nc}$ are fully assigned. Such a map, which will be dominated by local contacts, are minimally collapsible structures.

The white lines in Fig.(\ref{fig:pdbhist}a) show $k_\theta$ of chains of length $N$ with $N_\mathrm{nc}(N)$ contacts distributed in ways to maximize or minimize collapsibility. We estimated $N_\mathrm{nc}(N) \approx 0.6N^{\gamma}$, with $\gamma \approx 1.3$, from the fit of the proteins selected from the PDBSelect set ({ a fuller discussion is presented in Appendix A}). Since the lines are calculated for $N_\mathrm{nc}$ from the fit over the entire set, and not from $N_\mathrm{nc}$ for every protein, there are proteins below the minimal and above the maximal curve in Fig.(\ref{fig:pdbhist}a). For a given protein, with $N$ and $N_\mathrm{nc}$ defined by its PDB structure, $k_\theta$ for all possible arrangements of native contacts is largely in between the maximally  and minimally collapsible lines in Fig.(\ref{fig:pdbhist}a).  The majority of proteins in our set are closer to the maximal collapsible curves, suggesting that the unfolded proteins have evolved to be compact under native folding conditions. This theoretical prediction is in accord with our earlier studies which suggested that foldability is determined by both collapse and folding transitions \cite{Camacho93PNAS}, and more recently supported by experiments~\cite{Schuler16JACS}. 

\begin{figure}
	\includegraphics[width=0.9\textwidth]{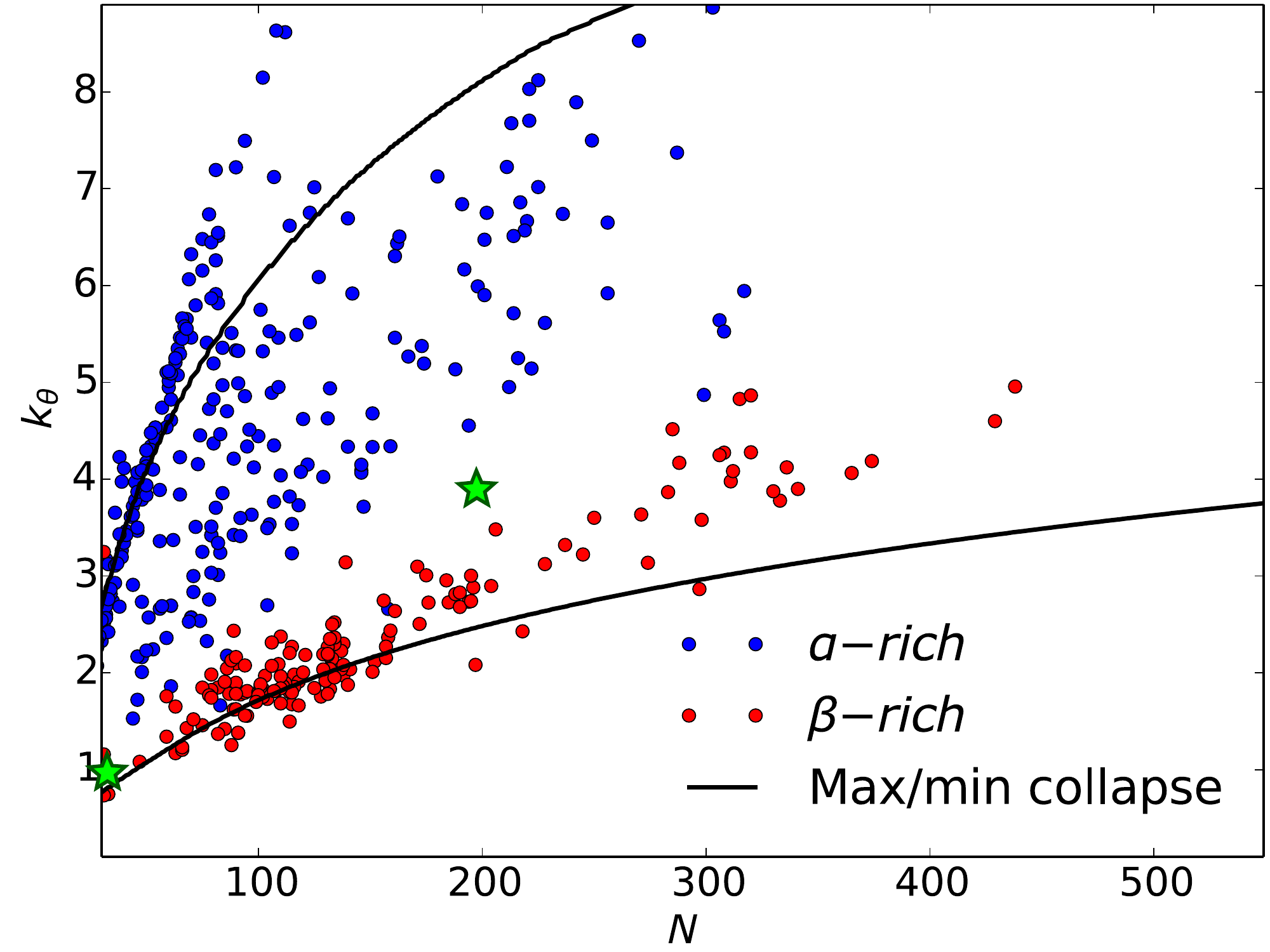}
	\caption{Dependence of $k_\theta$ on the secondary structure content of proteins. We display $k_\theta$ for $\alpha$-rich ($>90\%$) and $\beta$-rich ($>70\%$) proteins. Proteins that are predominantly $\alpha$-helical tend to be close to minimally collapsible (upper line), while $\beta$-rich proteins are close to  maximally collapsible curve (lower curve). The green stars are for RNA with the left one at small N corresponding to the\textit{Mouse Mammary Tumor Virus} (MMTV) pseudoknot (N=34) and the other is \textit{Azoarcus} ribozyme (N=196).}
	\label{fig:alphabeta}
\end{figure}

{\bf $\beta$-sheet rather than $\alpha$-helical proteins undergo larger compaction:} The weight function $W$ (Eq.~(\ref{weighting}) and Fig.(\ref{fig:pdbhist}b)) suggests that contacts in $\alpha$-helices ($|i-j|=4$) only make a small contribution to collapse. Contacts corresponding to the maximum of $W$ at $i-j\approx 30$ are typically found in loops and long antiparallel $\beta$-sheets. Fig.(\ref{fig:alphabeta}) shows a set of proteins with high $\alpha$-helix ($> 90\%$) and a set with high content of $\beta$-sheets ($>70\%$)~\cite{collapseURL}. The values of $k_\theta$ for the two sets are very distinct, so they barely overlap. We find that many of the $\alpha$-helical proteins lie on or above the curve of minimal collapsibility while the rest are closer to the maximal collapsibility. The smaller $\beta$-rich proteins lie on the curve of maximal collapsibility slightly diverging from it as the chain length grows. These results show that the extent of collapse of proteins that are mostly $\alpha $-helical is much less than those with predominantly $\beta$-sheet structures.

A note of caution is in order. The minimal collapsibility of most $\alpha$-helical proteins in the set may be a consequence of  some of them being transmembrane proteins, which do not fold in the same manner as globular proteins. Instead, the transmembrane $\alpha$-helices are inserted into the membrane by the translocon, one by one, as they are synthesized. Such proteins would not have the evolutionary pressure to be compact.

\begin{figure}
	\includegraphics[width=0.4\textwidth]{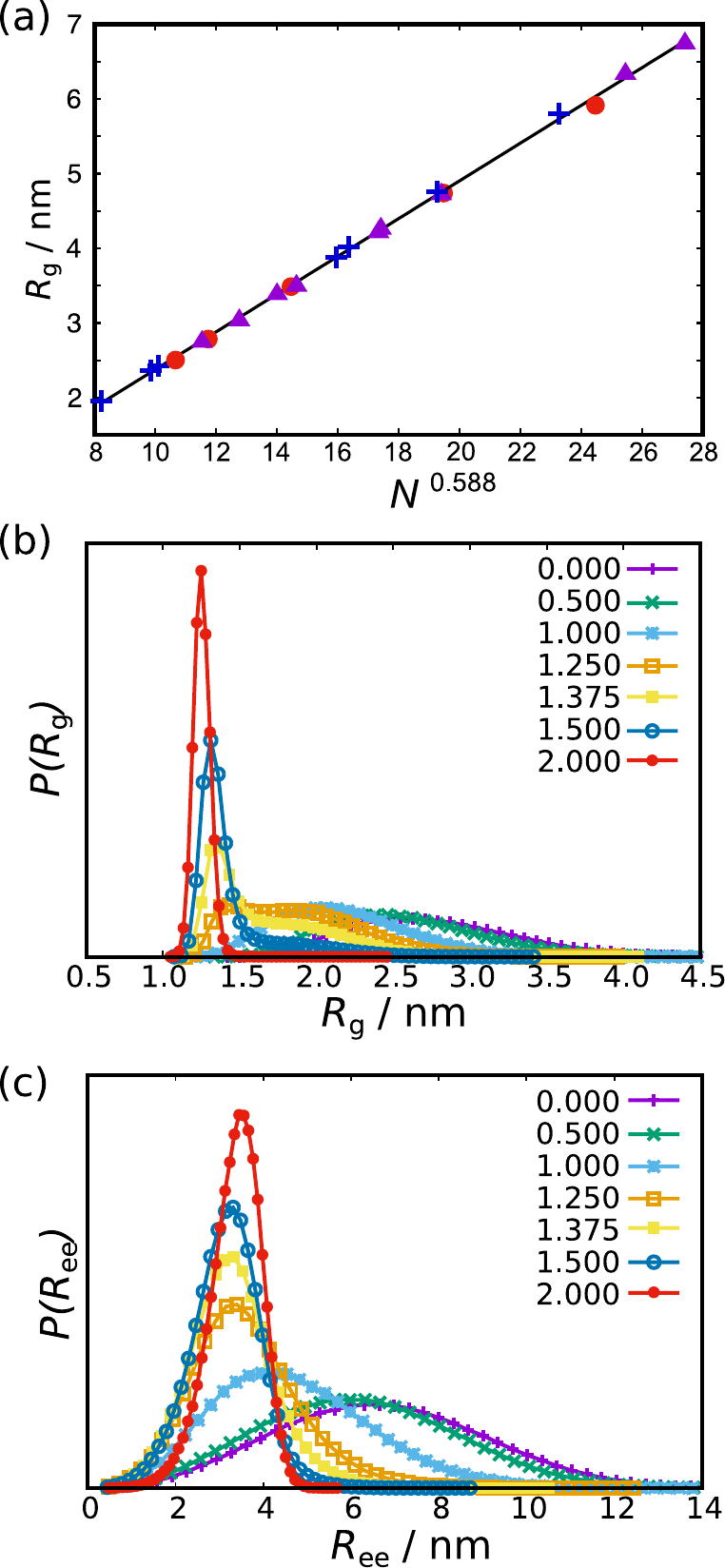}
	\caption{(a) Average $R_g$ at $k=0$ is plotted as a function of $N^{0.588}$ for the 21 proteins. The line is a fit, $R_g=0.25 N^{0.588} - 0.15$ (nm). (b) The probability distribution of the radius of gyration, $P(R_g)$ for different values of interaction strength $k$ for protein-L. As $k$ increases, the distribution becomes narrower. (c) Same as (a) except this panel shows end-to-end distribution $P(R_{ee})$ for different values of attractive strength $k$ for protein-L. The similarity between $P(R_{ee})$ and $P(R_g)$ shows that $R_{ee}$ also is a reasonable measure of compaction. }
	\label{fig:rg}
\end{figure}

\begin{figure}
\includegraphics[width=0.9\textwidth]{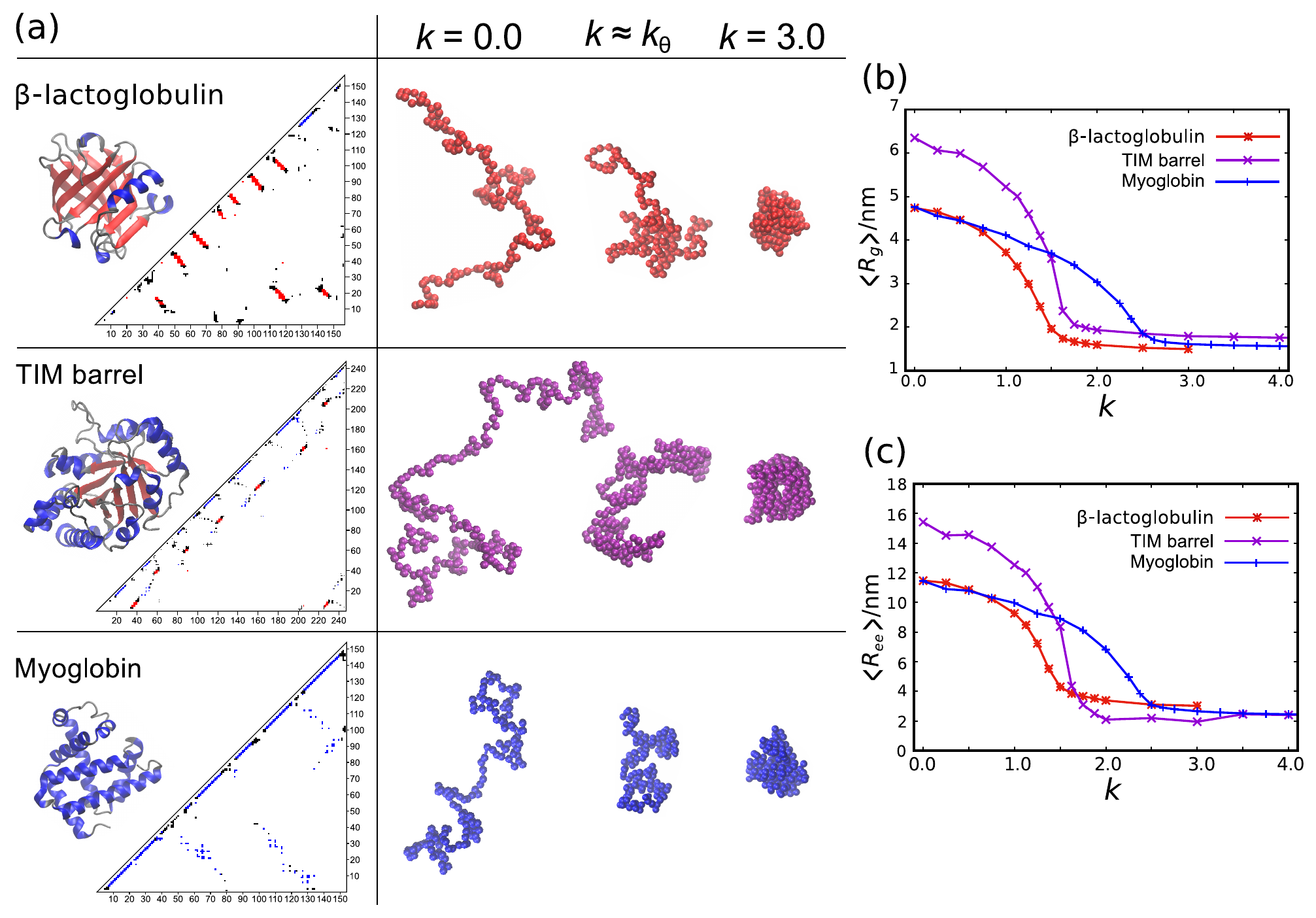}
\caption{\label{fig:simulationRg} Collapse transitions revealed by 
simulations for three representative proteins. (a) Contact maps and ribbon-diagram 
structures of three proteins, $\beta$-lactoglobulin (top), TIM
barrel (middle), and  myoglobin (bottom). Representative structures in the simulations are also shown for three values of $k$. (b) Average radius of gyration, $\left\langle R_{\textrm{g}}\right\rangle $,
monotonically decreases as $k$ increases. The three proteins with
different native topology have different $k_{\theta}$ values with myoglobin being less collapsible (larger $k_{\theta}$)
than $\beta$-lactoglobulin. (c)  Average end-end distance, $\left\langle R_{\textrm{ee}}\right\rangle $, also
monotonically decreases as $k$ increases although the changes in $R_{ee}$ are larger than in $R_g$. The middle panel shows snapshots from simulations at different $k$. The predicted conformation at $k \approx k_{\theta}$ is not random, supporting experiments showing persistent structures in the collapsed state of proteins. }
\end{figure}

\begin{figure}
\includegraphics[width=0.3\textwidth]{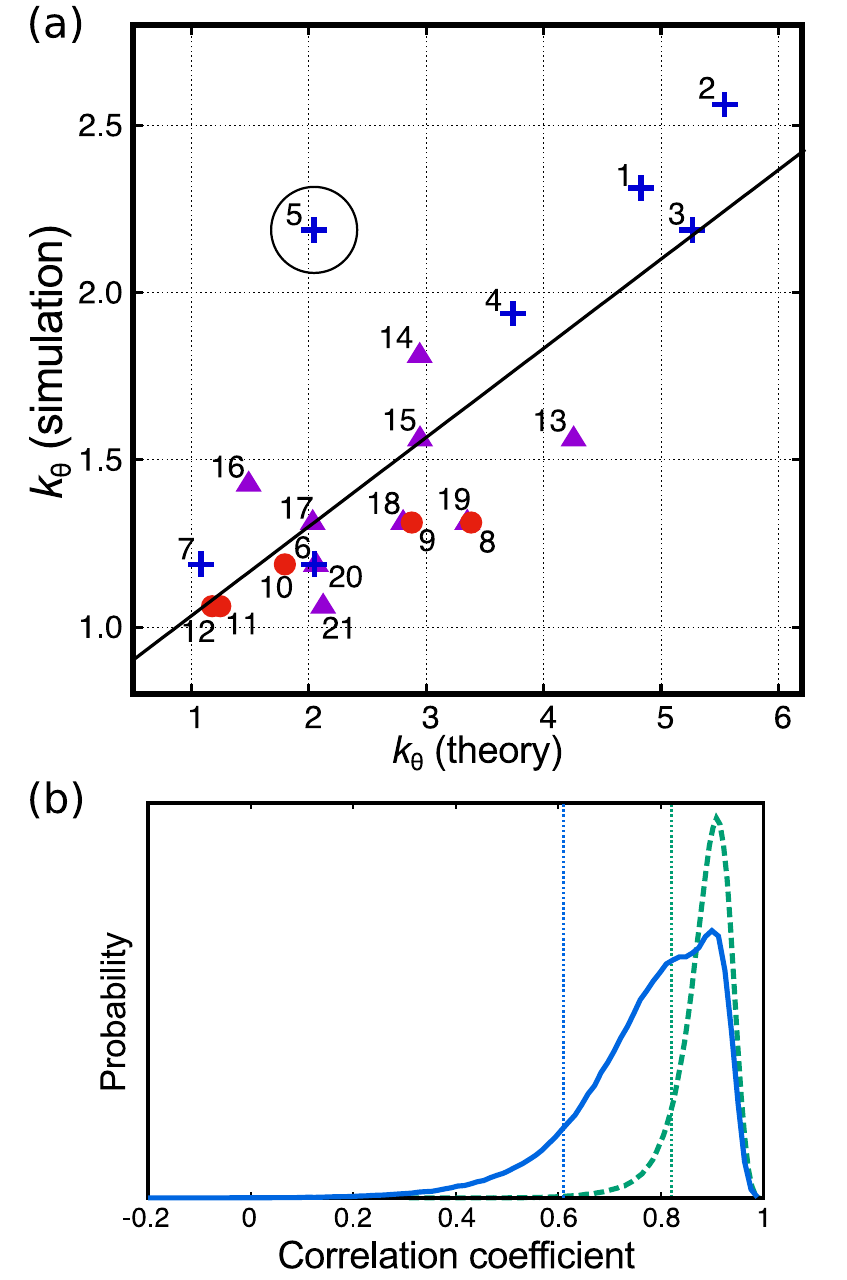}

\caption{\label{fig:correlation} (a) Correlation between simulation results and
theoretical predictions for $k_{\theta}$. The trends observed in simulations
are consistent with theoretical predictions. The horizontal axis is theoretical
$k_{\theta}$, and the vertical axis is $k_{\theta}$ value from simulations. In general, the theoretical $k_{\theta}$ values are larger than what is obtained in simulations (see the main text for an explanation), with the exception of protein labeled 5, a small all $\alpha$-helical B domain of protein A.
 In both theory and simulations, all-$\alpha$ proteins (blue crosses) have greater
$k_{\theta}$, and all-$\beta$ proteins (red circles) have smaller
$k_{\theta}$. The purple triangles are for proteins with $\alpha/\beta$
and $\alpha+\beta$ architecture. The linear-regression line for all
data points is shown in black and the Pearson correlation coefficient
is 0.79. Following is the complete list of 21 proteins with their PDB code and number of residues in parentheses. 1: Myoglobin (1mbo, 153); 2: Spectrin (3uun, 116); 3: Endonuclease III (2abk, 211); 4: BRD2 Bromodomain (5ibn, 111); 5: B domain of Protein A (1bdd, 51); 6: Villin headpiece (1vii, 36); 7: Homeodomain (1enh, 49); 8: GFP (1gfl, 230); 9: $\beta$-lactoglobulin (1beb, 156); 10: PDZ2 (1gm1, 94); 11: src SH3 (1srl, 56); 12: CspTm (1g6p, 66); 13: TIM Barrel (1r2r, 246); 14: Lysozyme (2lyz, 129); 15: CheY (3chy, 128); 16: Protein L (1K53, 64); 17: Barstar (1bta, 89); 18: RNase H (2rn2, 155); 19: Proteinase K (2id8, 279); 20: Ubiquitin (1ubq, 76); 21: Monellin (1iv9, 96).
(b) Population distribution of the correlation coefficient
estimated by the bootstrap analysis. The blue curve is generated for
the data set of all 21 proteins examined in the 
simulations. The mean of correlation coefficient is $\rho$ = 0.78
with $\rho_{90\%}>0.61$ (vertical dotted line) with 90\% confidence.
The distribution has two peaks indicating that there is at least one
outlier in the data set, which is the B domain of Protein
A. For remaining 20 proteins, the distribution has a single
peak (green broken line) with the mean 0.88 and $\rho_{90\%}>0.82$
(green dotted).}
\end{figure}

{\bf Comparison between theory and simulations:}
{ The major conclusions, summarized in Figs.(1-2), are based on an approximate
theory. In order to validate the theoretical predictions, we performed simulations for 21 proteins using  realistic models (see Appendix B for details) that capture the known characteristics of the unfolded states of proteins and the coil to globule transition.  

In accord with our theoretical predictions, $R_{\textrm{g}}$ decreases as $k$ increases. For $k=0$, corresponding to the maximally expanded state (high denaturant concentration) we expect that $R_{\textrm{g}} \approx a_D N^{0.588}$. A plot of $R_{\textrm{g}}$ versus $N^{0.588}$ is linear with a value of $a_D=0.25$ nm (Fig.\ref{fig:rg}a). Remarkably, this finding is in accord with the experimental fit showing $R_{\textrm{g}} \approx a_D N^{0.588}$ with $a_D = 0.2$ nm \cite{Kohn04PNAS}. The modest increase in the $a_D$, compared to the experimental fit, predicted here can be explained by noting that in real proteins there is residual structure even at high denaturant concentrations whereas in our model this is less probable.  The scaling shown in Fig. (\ref{fig:rg}a) shows that the model used in the simulations provides a realistic picture of the unfolded states. We emphasize that the parameters in the simulations were not adjusted to obtain the correct $R_{\textrm{g}}$ scaling or $a_D$.  

In Fig.~(\ref{fig:simulationRg}) we show the dependence of $R_{\textrm{g}}$ as a function of 
$k$ for three representative proteins along with their native and unfolded structures
and contact maps. The $\alpha$ helical protein myoglobin and the $\beta$-lactoglobulin with $\beta$ sheet architecture, have nearly the same number of
amino acids, $N\sim150$.  The
sizes of the two proteins are similar (Fig.\ref{fig:simulationRg}b) when $k$ is small ($k<0.5$) implying that the values of $R_{\textrm{g}}$ in the unfolded states are determined solely by $N$ (see Fig.\ref{fig:rg}a). For each protein, we identified 
 $k_{\theta}$ from simulations with the  $k$ value at which $\frac{dR_{g}}{dk}$ is a minimum.  
Using this method, we find that the $k_{\theta}$ value for  $\beta$-lactoglobulin is less than for
 myoglobin. This result is consistent
with the theoretical prediction,  demonstrating  that generically $\alpha$ proteins are less
collapsible than $\beta$ proteins. Interestingly, TIM barrel, an
$\alpha/\beta$ protein with larger chain length ($N=246$), collapses
at $k_{\theta}=1.6$, which is larger than $\beta$-lactoglobulin
but smaller than myoglobin (purple line in Fig.\ref{fig:simulationRg}b). These results are qualitatively consistent with theoretical predictions.

In Fig.~(\ref{fig:correlation}), we compare the predicted $k_{\theta}$ (Eq. (\ref{ktheta}))  and the values from
simulations.  The absolute values
of $k_{\theta}$ are different between simulations and theory because
we used entirely different   models to describe the coil to globule transition. The potential used in the theory, convenient for serving analytic expression for $k_{\theta}$, is far too soft to describe the structures of polypeptide chains. As a result the polypeptide chains explore small $R_{\textrm{g}}$ values without significant energetic penalty. Such unphysical conformations are prohibited in the realistic model used in the simulations. Consequently, we expect that the theoretical values of $k_{\theta}$ should differ from the values obtained in simulations.   
Despite the differences in the potentials used in theory and simulations, the trends in $k_{\theta}$ predicted using
  theory are the same as in  simulations. 
The Pearson correlation coefficient, $\rho=0.79$. Since we examined only
21 proteins in simulations, which is  fewer than theoretical predictions made for
2306 proteins, we analyzed the correlation data by the
bootstrap method to ascertain the statistical significance of $\rho$. The estimated
probability distribution of $\rho$ is shown in
Fig.~(\ref{fig:correlation}b). The mean of correlation coefficient is 0.78 and
$\rho_{90\%}>0.61$ with 90\% confidence. The distribution is bimodal
indicating that there is at least one outlier in the data set, 
which is likely to be the three helix bundle B domain of Protein A (labeled 5 in Fig.~(\ref{fig:correlation})). For 20 proteins excluding
Protein A, the distribution has a single peak (green broken line)
with the mean 0.88 and $\rho_{90\%}>0.82$ (green dotted line in Fig.~(\ref{fig:correlation})). From these results, we surmise that both theory and simulations qualitatively lead to the conclusion that proteins with $\beta$-sheet architecture are more collapsible than $\alpha$-helical  is structures, which is one of the major predictions of this work.

Given that the simulations describe the characteristics of the unfolded states, we show in Fig.(\ref{fig:rg}b) the variations in the probability distribution of $R_g$,
$P(R_g)$ for protein-L as a function of $k$. The broadest distribution, with $k=0$,
corresponds to the extended
chain. We find that $P(R_g)$ becomes narrower as the attractive strength ($k$)
increases. The continuous shift  to the compact state with
gradual increase in the attractive strength is consistent with experiments that the 
unfolded proteins collapse as the denaturant concentration decreases. Thus, generally $R_g$ of the $\bf{U_C}$ state is less than that of the
$\bf{U_D}$ state. The end-to-end distribution, $P(R_{ee})$, for
different values of values of $k$ in
Fig.(\ref{fig:rg}c) is broad at $k=0$ corresponding
to the unfolded protein. Average $R_{ee}$ decreases as attractive strength
increases and the distribution becomes narrower.  The results in
Fig.(\ref{fig:rg}) show that both $R_{ee}$, which can be inferred
using smFRET, and $R_g$ (measurable using SAXS), are smaller in the 
$\bf{U_C}$ state than the $\bf{U_D}$ state. However,
the extent of decrease is greater in $R_{ee}$ than $R_g$, an observation that has contributed to the smFRET-SAXS controversy.}

\begin{figure}
	\includegraphics[width=0.8\textwidth]{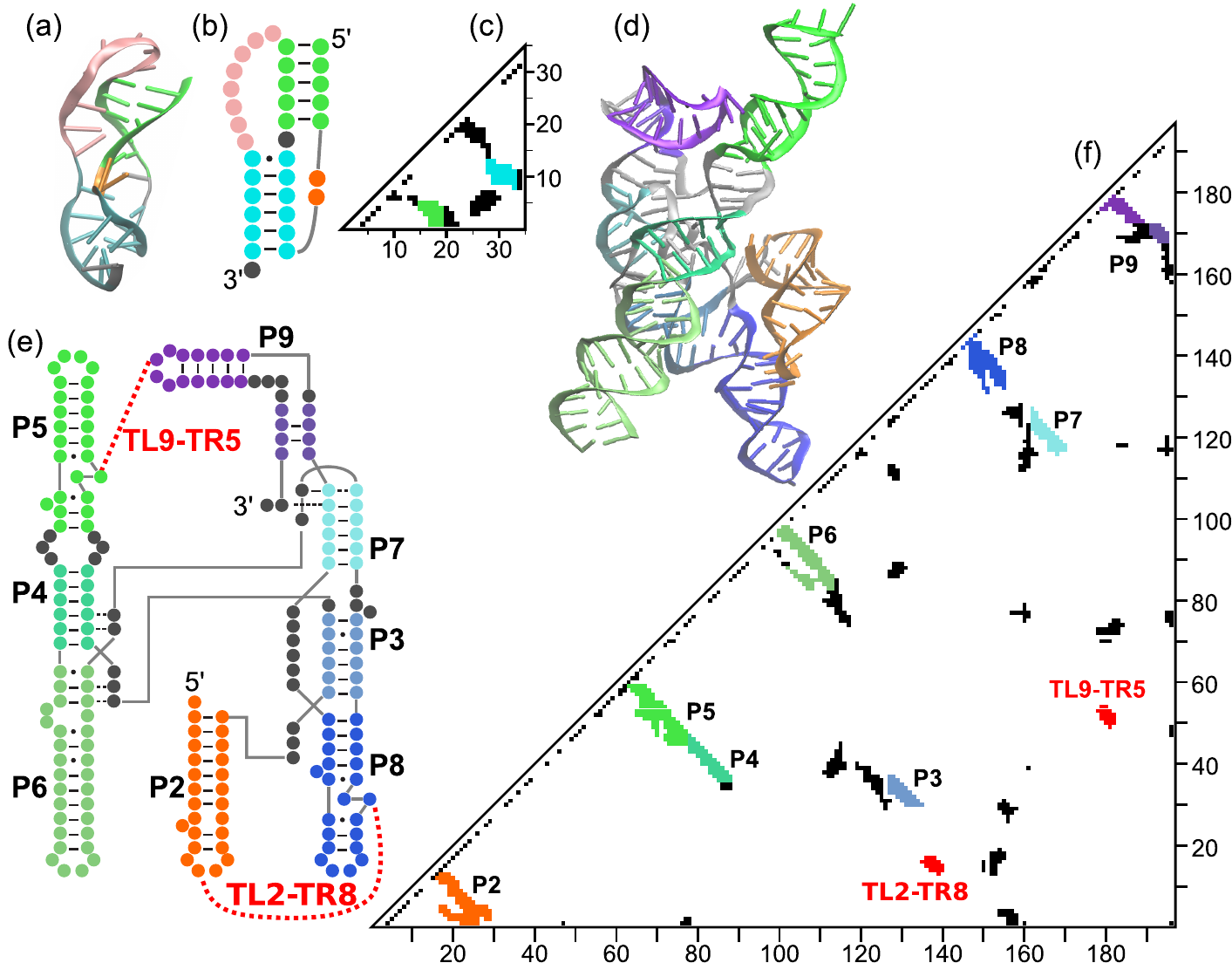}
	\caption{Native topologies of two RNA molecules, MMTV pseudoknot (a-c) and \textit{Azoarcus} ribozyme (d-f). Three-dimensional structures (a and d), secondary structures (b and e), and contact maps (c and f) are shown for each RNA. Colors are used to distinguish secondary structures. Contact pairs in RNA are defined as any nucleotide pair $i$ and $j$ ($|i-j|>2$) satisfying $R_{ij}<14\AA$, where $R_{ij}$ is the distance between centers of mass of the nucleotides \cite{Hyeon2007BJ}. MMTV has two stem basepairs (cyan and green in a-c), which contribute to non-local contacts (cyan and green in c).  \textit{Azoarcus} ribozyme has several hairpin basepairs (P2, P5, P6, P8 and P9) which can be seen in the vicinity of the diagonal in the contact map (f). There are also basepairs between nucleotides far along the sequence such as P3, P4 and P7, as well as tertiary interactions such as TL2-TR8 and TL9-TR5. These non-local contacts contribute to the collapsibility of the ribozyme.}
	\label{fig:cmap_rna}
\end{figure}

{\bf RNAs are compact:} There are major differences between how RNA and proteins fold \cite{Thirumalai05Biochem}. In contrast to the apparent controversy in proteins, it is well established that RNA molecules are compact\cite{Hyeon06JCP_2,Yoffe08PNAS,FangJCP11} at high ion concentrations or at low temperatures. Because our theory relies only on the knowledge  of contact map, used to assess collapsibility in {\it Azoarcus} ribozyme and MMTV pseudoknot to merely illustrate collapsibility of RNA (Fig.~(\ref{fig:cmap_rna})). The $k_{\theta}$ values (green stars in Fig.~(\ref{fig:alphabeta})) are close to the lower $\beta$-sheet line, indicating that these molecules must undergo compaction as they fold. This prediction from the theory is fully supported by both equilibrium and time-resolved SAXS experiments~\cite{Roh10JACS} on {\it Azoarcus} ribozyme. In this case ($N=196$) the changes are so large that even using low resolution experiments collapse is readily observed~\cite{Woodson12Cell}. We should emphasize that the size of different RNAs (for example viral, coding, non-coding) vary greatly. For a fixed length, single-stranded viral RNAs have evolved to be maximally compact, which is rationalized in terms of the density of branching. Although the sizes of the viral RNAs considered in \cite{Gopal14PLoSOne} are much longer than the {\it Azoarcus} ribozyme the notion that compaction is determined by the density of branching might be valid even when $N \sim200$. 

{{\bf Dependence of $k_\theta$ on the values of the cut-off:}

In order to ensure that the theoretical predictions  do not change qualitatively if the cutoff values are changed, we varied them over a reasonable range. The reason for our choice of $R_c$ is that in majority of folding simulations, using $C_{\alpha}$ representation of proteins, $R_c=0.8$ nm is typically used. Consider the variation of $k_{\theta}$ with $R_c$, the cut-off used to define contacts at a fixed $\sigma=0.63$ nm. As $R_c$ increases the number of contacts also increases. From Eq.~(\ref{ktheta}) it follows that $k_{\theta}$ should decrease, which is borne out in the results in Fig.(\ref{fig:appendix}a). Reassuringly, the trends are preserved. In particular, the prediction that $\beta$-sheet proteins are most collapsible is independent of $R_c$.  The trend that $\beta$-rich proteins are more collapsible than $\alpha$-rich proteins remains same irrespective of the $R_c$ values.  

Fig.(\ref{fig:appendix}b) shows the changes in $k_\theta $ for proteins as a function of $\sigma$ (contact distance) for fixed $R_c=0.8$ nm. The $k_\theta$ values decrease with increasing $\sigma$.  The predicted trend is independent of the precise value. It is worth emphasizing that the predictions based on simulations that the size of the proteins at $k_{\theta}$ is about (5-8)\% of the folded state was obtained using $\sigma=0.63$nm. This range is consistent with estimates based on experiments on a few proteins (see for example \cite{Denisov99NS&MB}). Higher values of $\sigma$ would give values of compact states of proteins that are less than the native state $R_g$. }

\section{Discussion}
We have shown that polymer chains with specific interactions, like proteins (but ones without a unique native state), become compact as the strength of the specific interaction changes. A clear implication is that the size of the ${\bf U_D}$ state should decrease continuously as $C$ decreases. In other words, the unfolded state under folding conditions is more compact than it is at high denaturant concentrations. Compaction is driven roughly by the same mechanism as the collapse transition in homopolymers in the sense that when the solvent quality is poor (below $C_m$) the size of the unfolded state decreases continuously.  When the set of specific interactions is taken from protein native contacts in the PDB, our theory shows that the values of $k_\theta$ are in the range expected for interaction between amino acids in proteins. This implies that collapsibility should be a universal feature of foldable proteins but the extent of compaction varies greatly depending on the architecture in the folded state. This is manifested in our finding that proteins dominated by $\beta$-sheets are more collapsible compared to those with $\alpha$-helical structures. 

{\bf Magnitude of $k_{\theta}$ and plausible route to multi-domain formation:}   The scaling of $k_{\theta}$ with $N$ allows us to provide arguments for the emergence of multi-domain proteins. In Eqs.~(\ref{self}) or 
(\ref{ktheta}) attractive ($\kappa$-) and repulsive ($v$-) terms have the same structure. The only difference in their scaling with $N$ is due to the difference in the sums (over all the monomers in the repulsive term and over native contacts in the attractive term). Double summation over all the monomers gives a factor of $N^2$ to the repulsive term. The summation over native contacts in the attractive term scales as $N_\mathrm{nc}$. Therefore, to compensate for the repulsion, $N_\mathrm{nc}$ should scale as $N^2$. However, for a given protein with a certain length $N$ and certain numbers of contacts, it is not clear how the denominator in Eq.~(\ref{ktheta}) scales with $N$. 
Empirically we find $N_\mathrm{nc}(N)$ dependence across a representative set of sequences scales as $N^{\gamma}$ with $\gamma$ at most $\approx$ 1.3 (Appendix A). Thus, it follows from Eq.~(\ref{ktheta}) that $k_\theta$ increases without bound as $N$ continues to increase. Because this is unphysical, it would imply that proteins whose lengths exceeds a threshold value $N_C$ cannot become maximally compact even at $C=0$. An instability must ensue when $N$ exceeds $N_C$. This argument in part explains why single domain proteins are relatively small\cite{Bussemaker97PRL}.

Scaling of $N_\mathrm{nc}$ as a power law in $N^{\gamma}$ means that as the protein size grows, the value of $k_\theta$
will deviate more and more from those found in globular proteins,
implying such proteins cannot be globally compact under physiologically
relevant conditions. However, such an instability is not a problem
because larger proteins typically consist of multiple domains. Thus,
if the protein does not show collapse as a whole, the individual
domains could fold independently, having lower values of
  $k_\theta$ for each domain of the multi-domain protein. It would be
interesting to know if the predicted onset of instability at $N_C$
provides a quantitative way to assess the mechanism of formation of
multi-domain proteins. Extension of the theory might yield interesting
patterns in the assembly of multi-domain proteins. For instance, one
can quantitatively ascertain if the N-terminal domains
of large proteins, which emerge from the ribosome
first, have higher collapsibility (lower $\kappa_\theta$)
than C-terminal domains.

{\bf SAXS-smFRET controversy resolved:} Our theory resolves, at least theoretically, the contradictory results using
SAXS and FRET experiments on compaction of small globular proteins. It
has been argued, based predominantly using SAXS experiments on protein-L ($N=72$)  that $R_g$ of $\bf{U_D}$ and $\bf{U_C}$ states are
virtually the same at denaturant concentrations that are less than
$C_m$ \cite{Yoo12JMB}. This conclusion is not only at variance with
SAXS experiments on other proteins but also with interpretation of
smFRET data on a number of proteins. The present work, surveying over
2300 proteins, shows that the compact state has to exist, engendered
by mechanisms that have much in common with homopolymer collapse. For
protein-L, the $k_\theta=1.7 k_BT$, a very typical value, is right on the
peak of the heat map in Fig.(\ref{fig:pdbhist}). We have previously
argued that because the change in $R_g$ between the $\bf{U_D}$ and
$\bf{U_C}$ states for small proteins is not large, high precision
experiments are needed to measure the predicted changes in $R_g$
between $\bf{U_C}$ and $\bf{U_D}$. For protein-L the change is less
than 10\%~\cite{Reddy16JACS}, making its detection in ensemble
experiments very difficult. Similar conclusions were reached in recent
experiments \cite{Schuler16JACS}. 
A clear message from our theory
is that, tempting as it may be, one cannot draw universal conclusions
about polypeptide compaction by performing experiments on just a few proteins. One has to survey a large number
of proteins with varying $N$ and native topology to quantitatively
assess the extent of compaction. Our theory provides a framework for
interpreting the results of such experiments.

\begin{figure}
	\includegraphics[height=0.6\textheight]{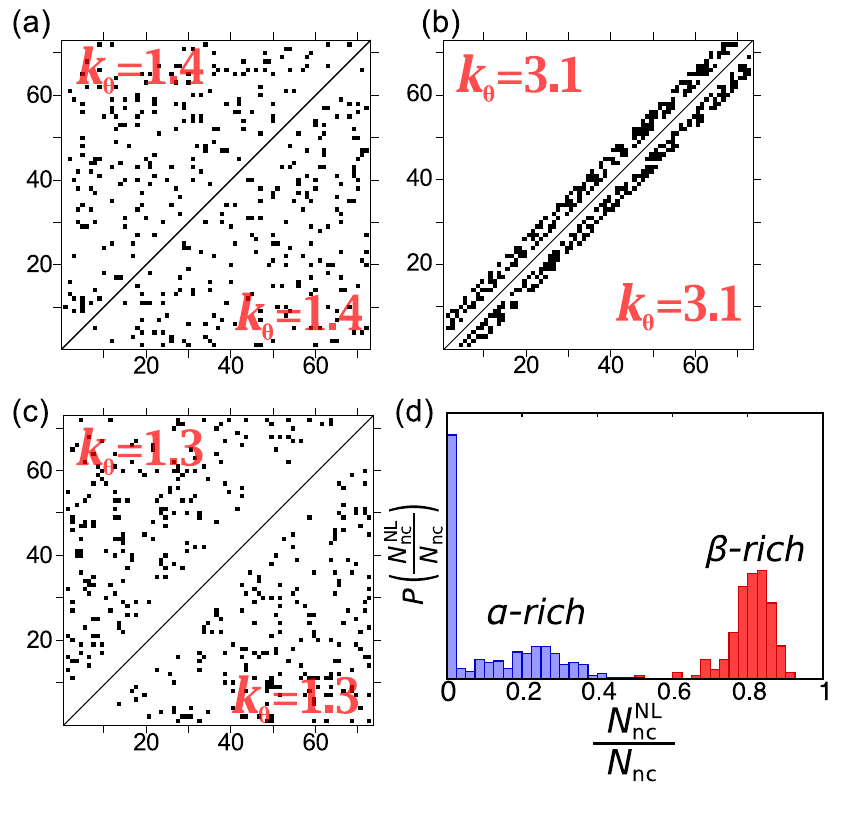}
	\caption{Collapsibility for synthetic contact maps. Two representative contact maps for each category are shown in the upper left and lower right of each square. Given the number of residues $N=72$ and total number of contacts $N_{nc}=185$ (same as protein-L), residue pairs $(i,j)$ are randomly chosen to satisfy the following conditions: (a) uniformly distributed, $|i-j|\geq3$; (b) local contacts only, $|i-j|\geq3$ and $|i-j|<8$;, and (c) non-local contacts only, $|i-j|\geq8$. The calculated values of $k_\theta$ are explicitly shown. The $k_\theta$ value for protein-L is $1.7 k_BT$.(d)
	Distribution of the fraction of non-local contacts in the 2306 proteins. For each protein, the fraction is calculated as the number of non-local contacts ($N^{\text{NL}}_{nc}$) divided by the total number of contacts ($N_{nc}$). A contact between residues $i$ and $j$ is ``non-local (NL)" if $|i-j|\geq8$. There is a clear separation in this distribution for proteins rich in $\alpha$ helices compared to those that are rich in $\beta$-sheets implying that the latter are more collapsible than the former.}

	\label{fig:random72}
\end{figure}

\begin{figure}
\includegraphics[height=0.6\textheight]{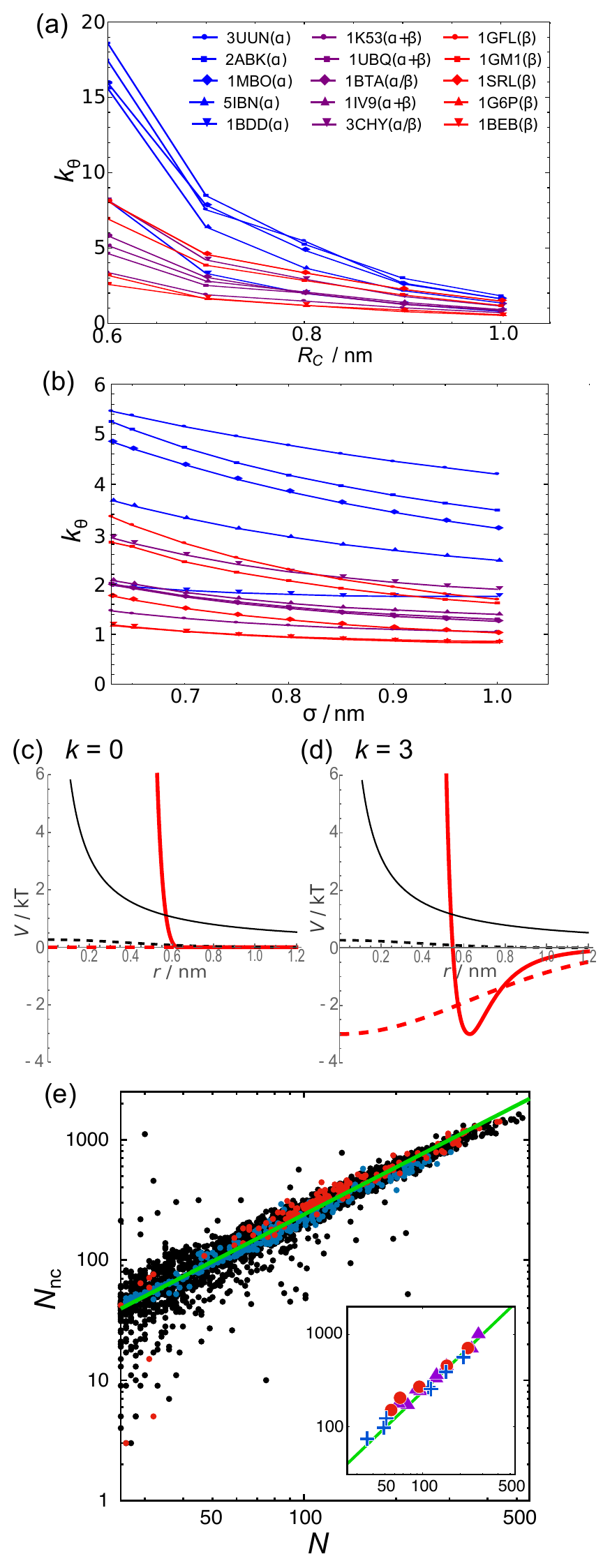}
\caption{\label{fig:appendix}(a) $k_\theta$ values for list of proteins with varying $R_c$ with fixed $\sigma=0.63$ nm. (b) $k_\theta$ values for list of proteins with varying $\sigma$ with fixed $R_c=0.8$ nm. (c and d) Comparison of potentials used in the theory and simulations with (c) $k$=0 and (d) $k$=3.  In the theory, Gaussian potentials are used for both non-specific repulsion (black broken line) and specific attraction (thick red-broken line).  In the simulations, potentials have hard core repulsion (black line) and WCA-type attraction (thick red line).  In both the theory and simulations, the depth of the attraction potential changes depending on values of $k$, whereas the repulsive part does not. The use of soft potentials in the theory results in larger $k_{\theta}$ than in simulations (Fig.~\ref{fig:correlation}). (e) The dependence of the number of contacts, $N_{nc}$, as a function of $N$ for the PDBselect proteins. $\alpha$-rich and $\beta$-rich proteins are colored in blue and red, respectively. The green line is a fit using $N_{nc}=0.6 N^{\gamma}$ with $\gamma=1.3$.  A plot for 21 proteins used in the simulations is shown in the inset.}
\end{figure}

{\bf Random contact maps, local and non-local contacts:} In order to differentiate collapsibility between evolved and random proteins, we created twelve random contact maps keeping the total number of contacts the same as in protein-L (see Fig.(\ref{fig:random72}) for examples). For each of these pseudo-proteins we calculated $k_{\theta}$ using Eq.~(\ref{ktheta}). We find that for all the random contact maps the $k_{\theta}$ values are less than for protein-L, implying that the propensity of the pseudo-proteins to become compact is greater than for the wild type. 
This finding is in accord with studies based on homopolymer and heteropolymer collapse with random crosslinks. These studies showed that the polymer undergoes a collapse transition as the density of crosslinks is increased~\cite{Bryngelson96PRL,Kantor96PRL,Zwanzing97JCP}. Of particular note is the demonstration by Camacho and Schanke~\cite{Camacho97EPL}, who showed using exact enumeration of random heteropolymers and scaling arrangements that the collapse can be either a first or second order transition depending on the fraction of hydrophobic residues \cite{Camacho97EPL}. 

Some time ago Abkevich et al.~\cite{AbkevichJMB95} showed, using Monte Carlo simulations of protein-like lattice polymers, that the folding transition in  proteins with predominantly non-local contacts  was first order like, which is not the case for proteins in which local contacts dominate. In light of this finding, it is interesting to examine how compaction is affected by local and non-local contacts. We created for $N$=72  (protein-L) a contact map with 185 (same number as with WT protein-L), predominantly local contacts (Fig.(\ref{fig:random72}b)). The values of $k_\theta$ for these pseudo-proteins is considerably larger than for the WT, implying that proteins dominated by local contacts are minimally collapsible. We repeated the exercise by creating contact maps with predominantly non-local contacts (Fig.(\ref{fig:random72}c)). Interestingly, $k_\theta$ values in this case are significantly less than for the WT. This finding explains why in proteins with varied $\alpha/\beta$ topology there is a balance between the number of local and non-local contacts. Such a balance  is needed to achieve native state stability and speed of folding~\cite{AbkevichJMB95} with polypeptide compaction playing an integral part \cite{Camacho93PNAS}.

Based on these findings we conclude that $R_g$ of the unfolded states of proteins dominated by non-local contacts must undergo greater compaction compared to those with that have mostly local contacts. The results in Fig.~(\ref{fig:alphabeta}) also show that proteins rich in $\beta$-sheet are more collapsible than predominantly $\alpha$-helical proteins. It follows that $\beta$-sheet proteins must have a larger fraction of non-local contacts than proteins rich in $\alpha$-helices. In Fig. (\ref{fig:random72}d) we plot the distribution of the fraction of non-local contacts for the 2306 proteins. Interestingly, there is a clear separation in the distribution of non-local contacts between $\alpha$-helical rich and $\beta$-sheet rich proteins. The latter have substantial fraction of non-local contacts which readily explains the findings in Fig. (\ref{fig:random72}c) and the predictions in Fig. (\ref{fig:alphabeta}).

\section{Conclusions}
We have created a theory to assess collapsibility of proteins { using a combination of analytical modeling and simulations.} The major implications of the theory are the following. (i) Because single domain proteins are small, the changes in the radius
of gyration of the unfolded states as the denaturant concentration is lowered are often small. Thus, 
it has been difficult to detect the $R_{\textrm{g}}$ changes using SAXS experiments in a couple of proteins, raising the 
question if unfolded polypeptide chains become compact below $C_m$.
Here, we have solved this long-standing problem showing
that the unfolded states of single-domain proteins do become compact
as the denaturant concentration decreases, sharing much in common
with the physical mechanisms governing homopolymer
collapse. By adopting concepts from polymer physics, and using the
contact maps that reflect the topology of the native states, we
established that proteins are collapsible. { Simulations using models that describe the unfolded states of proteins reasonably well further confirm the conclusions based on theory.} (ii) Based on a survey of over
two thousand proteins we surmise that there is evolutionary pressure
for collapsibility is universal although the extent of collapse can vary greatly, because this ensures that the propensity to
aggregate is minimized even if environmental fluctuations under
cellular conditions transiently populate unfolded
states. Two factors contribute to aggregation. First, the rate of dimer formation by diffusion controlled reaction would be enhanced if a pair of $\bf{U_D}$ rather than $\bf{U_C}$ molecules collided due cellular stress because the contact radius in the former would be greater than in the latter. Second, the fraction of exposed hydrophobic resides in  $\bf{U_D}$  is much greater than in $\bf{U_C}$, thus greatly increasing the probability of aggregation. The second factor is likely to be more important than the first. Consequently, transient population of  $\bf{U_C}$ due to cellular stress minimizes the probability of aggregation.  (iii)  We have also shown that the position of the
residues forming the native contact greatly influences the
collapsibility of $\beta$ sheet proteins (containing a
number of non-local contacts showing greater compaction than $\alpha$
helical proteins, which are typically stabilized by local contacts.

Our theory also shows that most RNAs may have evolved to be compact in their natural environments. Although the evolutionary pressure to be compact is likely to be substantial for viral RNAs \cite{FangJCP11,Yoffe08PNAS, Gopal14PLoSOne, TubianaBJ15}, it is apparent that even non-coding RNAs are also likely to be almost maximally compact in their natural environments. Our theory suggests that, to a large extent, collapsibility of RNA is similar to proteins with $\beta$-sheet structures. Both classes of biological macromolecules are stabilized by non-local contacts. Interestingly, it has been argued that the need to be compact (``Compaction selection hypothesis'' \cite{TubianaBJ15}) could be a major determinant for evolved biopolymers to have minimum energy compact structures as their ground states.


\appendix
{
\section{{}}
{\bf Collapse of homopolymers:}
The theory described for protein collapse resulting in Eq. (\ref{ktheta}) is general and applicable to the collapse of homopolymers as well. We show in this Appendix that the  ES formalism can be used to derive the scaling of $k_{\theta}$ with $N$, the number of monomers. 

Consider a homopolymer with the following Hamiltonian:
\begin{equation}\label{hamiltonianH}
\mathcal{H}=\frac{3k_B T}{2 a_0^2} \int\limits_0^N  \left(\frac{\partial {\bf r}}{\partial s}\right)^2 ds + k_B T ~ {V_H}({\bf r}(s)),
\end{equation}
where ${\bf r}(s)$ is the position of the monomer $s$, and $a_0$ is the monomer size.  The first term in Eq.(\ref{hamiltonianH}) accounts for chain connectivity, and the second term represents volume interactions and favorable interactions between monomers, given by ${V}_H({\bf r}(s))$,  
\begin{equation}\label{HamiltVH}
{V_H}({\bf r}(s))=\frac{v}{(2\pi a_0^2)^{3/2}}\sum\limits_{s=0}^{N}  \sum\limits_{s'=0}^{N}  
e^{-\frac{({\bf r}(s)-{\bf r}(s'))^2}{2a_0		^2}}-
\frac{\kappa}{(2\pi \sigma^2)^{3/2}}\sum\limits_{s=0}^{N}  \sum\limits_{s'=0}^{N}  e^{-\frac{({\bf r}(s)-{\bf r}(s'))^2}{2\sigma^2}}
\end{equation}

The form of ${V_H}({\bf r}(s))$ is exactly the same as in Eq. \ref{HamiltV} except in the above equation all monomers interact favorably as long as self-avoidance is not violated whereas in Eq. (\ref{HamiltV}) attractive interactions depend on the topology of the protein. The first (second) term in Eq.~(\ref{HamiltVH}) describes  non-specific excluded volume (attractive) interactions. Thus, the model in Eq.~(\ref{hamiltonianH}) describes the behavior in good solvents ($k=0$) as well as the transition point at which there is a transition to the collapsed state. For the excluded volume repulsion, the range of interactions is on the order of the size of the monomer $a_0$ and for attractive interactions, the range is $\sigma$.
In good solvents, with $v > 0$, the polymer swells with $R_g \sim a N^{\nu}$ $(\nu \approx 0.6)$. In poor solvents ($v < 0$),  the polymer undergoes a coil-globule transition with $R_g \sim a N^\nu$ ($\nu = \frac{1}{3}$). These are the well-known Flory laws.  

Following the ES method described in the main text, we arrive at the self-consistent equation for $a$ for the homopolymer chain,
\begin{eqnarray}\label{selfH}
  \frac{1}{a_0^2}&-&\frac{1}{a^2}=v\frac{ (\frac{3}{2})^{5/2}(\frac{\pi}{2})^{3/2}}{ (a^2)^{5/2}N^{3/2}\left(\sum\limits_{n=1}^N \frac{1}{n^2}\right)}
  \sum\limits_{s=0}^N  \sum\limits_{s'=0}^N   \frac{\sum\limits_{n=1}^{N}\frac{1-\cos[n \pi(s-s')/N]}{n^4}}
{\left(\sum\limits_{n=1}^{N}\frac{1-\cos[n \pi(s-s')/N]}{n^2}+\frac{3 \pi^2a_0^2 }{2a^2 N}\right)^{5/2}}\\ \nonumber
  &-& \kappa\frac{ (\frac{3}{2})^{5/2}(\frac{\pi}{2})^{3/2}}{(a^2)^{5/2}N^{3/2}\left(\sum\limits_{n=1}^N \frac{1}{n^2}\right)}
\sum\limits_{s=0}^N  \sum\limits_{s'=0}^N   \frac{\sum\limits_{n=1}^{N}\frac{1-\cos[n \pi(s-s')/N]}{n^4}}
{\left(\sum\limits_{n=1}^{N}\frac{1-\cos[n \pi(s-s')/N]}{n^2}+\frac{3 \pi^2\sigma^2 }{2a^2 N}\right)^{5/2}}.\\ \nonumber
\end{eqnarray}

To obtain an expression for the $\theta$-point  we derive the condition for homopolymer collapse instead of solving the complicated Eq.~(\ref{selfH}) numerically. The volume interactions are on the right hand side of Eq.~(\ref{selfH}). At the $\theta$-point, the $v$-term  should exactly balance the $\kappa$-term arising from attractive interaction between the monomers. Since at the $\theta$-point the chain is ideal with $a=a_0$, we can substitute this value for $a$ in the sums in the denominators of the $v$- and $\kappa$-terms, to obtain an expression for $\kappa_\theta$.
Thus, from Eq.~(\ref{selfH}), the specific interaction strength at which two-body repulsion ($v$-term) equals two-body attraction ($\kappa$-term) is:
\begin{equation}
	\kappa_\theta=\frac{4}{3}\pi a_0^3\frac{\sum\limits_{s=0}^N  \sum\limits_{s'=0}^N   \frac{\sum\limits_{n=1}^{N}\frac{1-\cos[n \pi(s-s')/N]}{n^4}}
	{\left(\sum\limits_{n=1}^{N}\frac{1-\cos[n \pi(s-s')/N]}{n^2}+\frac{3 \pi^2 }{2 N}\right)^{5/2}}}{ \sum\limits_{s=0}^N  \sum\limits_{s'=0}^N   \frac{\sum\limits_{n=1}^{N}\frac{1-\cos[n \pi(s-s')/N]}{n^4}}
	{\left(\sum\limits_{n=1}^{N}\frac{1-\cos[n \pi(s-s')/N]}{n^2}+\frac{3 \pi^2\sigma^2 }{2a_0^2 N}\right)^{5/2}}}.
\label{kthetaH}
\end{equation}
The expression for $k_\theta$ in Eq. (\ref{kthetaH}) for homopolymers differs from $k_\theta $ (Eq. (\ref{ktheta})) for proteins only by the term in the denominator. The sum over specific interactions for proteins is replaced by the non-specific interaction in Eq. (\ref{kthetaH}). It can be shown that the $N$ dependence is the same in both the numerator and denominator in Eq. (\ref{kthetaH}). Therefore, to leading order in $\mathcal{W}$, $k_\theta $ is independent of $N$ for a homopolymer. 
 
In order to derive the scaling of $k_{\theta}$ with  $N$,  we need to analyze the corrections arising from second order in $\mathcal{W}$.
To second order in $\mathcal{W}$, the radius of gyration is,
\begin{eqnarray}\label{rg}
\langle R_g^2\rangle=\frac{1}{N} \int\limits_0^N \langle{\bf r}^2(s)\rangle ds &=& \frac{1}{N} \int\limits_0^N [\langle{\bf r}^2(s)\rangle_v + \langle{\bf r}^2(s)\rangle_v \langle\mathcal{W}\rangle_v -\langle{\bf r}^2(s)\mathcal{W}\rangle_v \\ \nonumber
&&\frac{1}{2}(\langle{\bf r}^2(s)\mathcal{W}^2\rangle_v-\langle{\bf r}^2(s)\rangle_v \langle\mathcal{W}^2\rangle_v)
] ds,
\end{eqnarray}
In the expression $\frac{1}{2}(\langle{\bf r}^2(s)\mathcal{W}^2\rangle_v-\langle{\bf r}^2(s)\rangle_v \langle\mathcal{W}^2\rangle_v)$, only the  $\frac{1}{2}(\langle{\bf r}^2(s)\mathcal{W}_2^2\rangle_v-\langle{\bf r}^2(s)\rangle_v \langle\mathcal{W}_2^2\rangle_v)$  contribute to $k_\theta$. Here, $\mathcal{W}_1$ is the same as Eq. (5), and $\mathcal{W}_2$ is given by Eq. \ref{HamiltVH}. The terms associated with $\mathcal{W}_1$ are zero at the $\theta $-transition point. By counting the powers of $N$ it follows  that $\langle{\bf r}^2(s)\mathcal{W}_2^2\rangle_v$ scales as $\frac{1}{N^7}$ and $\langle{\bf r}^2(s)\rangle_v \langle\mathcal{W}_2^2\rangle_v$ scales as $\frac{1}{N^5}$. 
Hence, at the $\theta$-point, we find that $k_\theta$ satisfies the following quadratic equation,
\begin{equation}
k_\theta^2+(2N^{1/2}-2v)k_\theta -v(2N^{1/2}-v)=0 ~ \Longrightarrow k_\theta \sim v(1-\frac{v}{2} N^{-1/2}).
\end{equation}
in the large $N$ limit. 
The scaling law for $k_\theta$ ($\propto T_{\theta}$) obtained first by Flory~\cite{Flory}, was confirmed using simulations much later~\cite{Wilding96JCP}. To our knowledge this is the first microscopic derivation of the result. Thus, our general formalism can be applied to  describe collapse of homopolymers as well as proteins and RNA.

{\bf Proteins:} The results for homopolymers given above may be extended to obtain the $N$ dependence of $k_{\theta}$ for proteins. By considering the second order correction to the radius of gyration, we obtain the following quadratic equation for $k_\theta$,
\begin{equation}
k_\theta^2 N^{6(1-\gamma)}+(2N^{(7/2-3\gamma)}-2 v N^{3(1-\gamma)})k_\theta -v(2N^{1/2}-v)=0 ~ \Longrightarrow k_\theta \sim N^{3(\gamma-1)}.
\end{equation}
In deriving the above equation we assume that total number of contacts $N_{nc}\sim N^\gamma$. A plot of $N_{nc}$ as a function of $N$ (Fig.~(\ref{fig:appendix}e)) for the PDBselect proteins confirms that this is indeed the case.  For $\gamma =1.3$, $k_\theta \sim N^{0.9}$, which  shows that larger proteins are less collapsible than smaller ones,  implying that when $N$ exceeds a critical value they are likely to form multi-domain structures.  Comparison of Eqs. (A6) and (A7) shows that collapsibility in proteins and homopolymers differs dramatically.  For homopolymers the coil-to-globule transition occurs at a finite temperature. The sharpness of the transition increases as $N$ increases. In sharp contrast, the growth of $k_{\theta}$ with $N$ for proteins (Eq. (A7)) implies that larger proteins must organize themselves into domains with individual domains forming compact structures.

\section{Simulations}

The theoretical results were obtained using a set of approximations, whose validity need to be confirmed using simulations. The purpose of these simulations is to show that the predicted theoretical values of $k_{\theta}$ correlate well with simulation results. We performed Langevin dynamics simulations for 21 globule proteins
(Fig.~(\ref{fig:correlation})). The set includes both all-$\alpha$
and all-$\beta$ proteins as well as $\alpha+\beta$ and $\alpha/\beta$
proteins according to Structural Classification Of Proteins (SCOP).

The simple form (sum of Gaussians) of the interaction energy in Eq.~(\ref{HamiltV}) was devised in order to obtain analytic expression for $k_{\theta}$ so that collapsibility of two thousand or more proteins could be easily analyzed. The potential in Eq.~(\ref{HamiltV}) has no hard core, which is physically not realistic. Because of the soft interactions it is clear that the theoretical values of $k_{\theta}$ have to be an upper bound. In order to firmly establish the qualitative predictions obtained using theory we use a realistic interaction energy in the simulations.  The potential function in the simulations is,
\begin{equation}\label{VSIM}
V_S=\frac{3k_{B}T}{2a_{0}}\sum_{i=1}^{N-1}r_{i,i+1}^{2}+\sum_{i,j\notin con}\varepsilon_{v}\left(\frac{a_{0}}{r_{ij}}\right)^{12}+\sum_{i,j\in con}\Phi_{\textrm{WCA}}\left(\varepsilon_{v},\varepsilon_{k}\right),
\end{equation}
where
\begin{equation}
\Phi_{\textrm{WCA}}\left(\varepsilon_{v},\varepsilon_{k}\right)=\begin{cases}
\varepsilon_{v}\left[\left(\frac{\sigma}{r}\right)^{12}-2\left(\frac{\sigma}{r}\right)^{6}+1\right]-\varepsilon_{k} & (r<\sigma)\\
\varepsilon_{k}\left[\left(\frac{\sigma}{r}\right)^{12}-2\left(\frac{\sigma}{r}\right)^{6}\right] & (r\geq\sigma)
\end{cases}.
\end{equation}

The first term, describing chain connectivity, the is discrete version of the first term in Eq.~(\ref{hamiltonian}) with $a_{0}=0.38$ nm. The second term accounts for excluded volume interactions used for any pair of residues not included in the contact map.  We chose $\varepsilon_{v}=1.0\text{ kcal/mol}$
so that monomer particles do not overlap with each other. In this crucial respect,
the potential function is drastically  different from the interaction potential
used in the theory, in which the Gaussian-type soft core potential
was used in order to solve the problem analytically. 

The summation in 
the last term in Eq.~(\ref{VSIM}) runs over all pairs in the contact map. The potential,
$\Phi_{\textrm{WCA}}$, is the Weeks-Chandler-Andersen potential~\cite{Weeks71},
a variant of Lenard-Jones potential, consisting of well-separated
repulsive and attractive terms (Fig.~\ref{fig:appendix}(c), (d)). This is necessary
in order to vary the strength of the attraction potential without affecting
the repulsive interactions.  The coefficient
of the attractive term is $\varepsilon_{k}=k\cdot k_{\textrm{B}}T$.
We varied $k$ between 0.0 and 5.0 to find the collapse-transition point,
$k=k_{\theta}$. The contact distance is the same as in the theory, $\sigma=0.63$ nm.

For each protein  and $k$ value, we generated 100 independent simulation trajectories.  Initial conformations were generated in a preliminary simulation
at high temperature $T=400$ K with $k=0$. Each production run at
$T=300$ K lasts for $10^{8}$ steps. We discarded the first $2\times10^{7}$ steps
in analyzing the data. Conformations are sampled every
$10^{4}$ steps. In total, $8\times10^{5}$ conformations were sampled
to calculate the average radius of gyration, $\left\langle R_{\textrm{g}}\right\rangle $
for each $k$.
}

\section*{Acknowledgements:}
This work was supported by a grant from the National Science Foundation (CHE 16-36424). We acknowledge the Texas Advanced Computing Center (TACC) at The University of Texas at Austin for providing resources for the simulations.

\bibliographystyle{unsrt}

\clearpage

\end{document}